\documentclass[%
 reprint,
superscriptaddress,
amsmath,amssymb,
aps,
prb,
]{revtex4-1}
\usepackage{placeins}
\usepackage{float}
\usepackage{mathtools}
\usepackage{siunitx }
\usepackage{chemformula}
\usepackage{braket}
\usepackage{xparse}
\usepackage{amsmath}
\usepackage{graphicx}% Include figure files
\usepackage{dcolumn}% Align table columns on decimal point
\usepackage{bm}% bold math
\usepackage{hyperref}% add hypertext capabilities
\usepackage{color}

\begin{document}

\preprint{APS/123-QED}

\title{Strength of Correlations in a Silver Based Cuprate Analogue}% Force line breaks with \\

\author{Riccardo Piombo}
%\email[Email address: ]{riccardo.piombo@uniroma1.it}
\affiliation{Dipartimento di Fisica, Universit\`a  di Roma ``La Sapienza'', 00185 Rome, Italy}

\author{Daniel Jezierski}
\affiliation{Center of New Technologies, University of Warsaw, 02089 Warsaw, Poland}

\author{Henrique Perin Martins}%
\affiliation{Departamento de Fisica, Universidade Federal do Paran\'a,\\
Caixa Postal 19044, 81531-990, Curitiba, PR, Brazil}%

\author{Tomasz Jaro\'n}
\affiliation{Center of New Technologies, University of Warsaw, 02089 Warsaw, Poland}

\author{Maria N. Gastiasoro}
\affiliation{Institute for Complex Systems (ISC), Consiglio Nazionale delle Ricerche, Dipartimento di Fisica, Universit\`a di Roma ``La Sapienza'', 00185 Rome, Italy}

\author{Paolo Barone}
\affiliation{{Institute for superconductors, oxides and other innovative materials (SPIN-CNR), Area della Ricerca di Tor Vergata, Via del Fosso del Cavaliere 100, 00133 Rome, Italy}}%

\author{Kamil Tok\'ar}
\affiliation{
Advanced Technologies Research Institute, Faculty of Materials Science and Technology in Trnava, Slovak University of Technology in Bratislava, J. Bottu 25, 917 24 Bratislava, Trnava, Slovakia}
\affiliation{
Institute of Physics, Slovak Academy of Sciences, D\'ubravsk\'a cesta 9, 845 11 Bratislava, Slovakia}

\author{Przemys\l aw Piekarz}
\affiliation{Institute of Nuclear Physics, Polish Academy of Sciences, Radzikowskiego 152, 31342 Krak\'ow, Poland}

\author{Mariana Derzsi}
\affiliation{
Advanced Technologies Research Institute, Faculty of Materials Science and Technology in Trnava, Slovak University of Technology in Bratislava, J. Bottu 25, 917 24 Bratislava, Trnava, Slovakia}

\author{Zoran Mazej}
\affiliation{Jo\u{z}ef Stefan Institute, Department of Inorganic Chemistry and Technology, Jamova cesta 39, 1000 Ljubljana, Slovenia}

\author{Miguel Abbate}
\affiliation{Departamento de Fisica, Universidade Federal do Paran\'a¡,\\
Caixa Postal 19044, 81531-990, Curitiba, PR, Brazil}%

\author{Wojciech Grochala}
\email[Email address: ]{w.grochala@cent.uw.edu.pl}
\affiliation{Center of New Technologies, University of Warsaw, 02089 Warsaw, Poland}%

\author{Jos\'e Lorenzana}
\email[Email address: ]{jose.lorenzana@cnr.it}
\affiliation{Institute for Complex Systems (ISC), Consiglio Nazionale delle Ricerche, Dipartimento di Fisica, Universit\`a di Roma ``La Sapienza'', 00185 Rome, Italy}

\begin{abstract}
\ch{AgF2} has been proposed as a cuprate analogue which requires strong correlation and marked covalence. On the other hand, fluorides are usually quite ionic and $4d$ transition metals tend to be less correlated than their $3d$ counterparts which calls for further scrutiny. We combine %core level, 
valence band photoemission and Auger-Meitner  spectroscopy of AgF and \ch{AgF2} together with computations in small clusters 
to estimate values of the \ch{Ag} $4d$ Coulomb interaction $U_{4d}$ and charge-transfer energy $\Delta_{pd}$.
%We find $\Delta_{pd}=3.8\pm 0.3$ eV and   $U_{4d}=7.7\pm 0.5$ eV.
Based on these values \ch{AgF2} can be classified as a charge-transfer correlated insulator according to the Zaanen-Sawatzky-Allen classification scheme. Thus, we confirm that the material is a cuprate analogue from the point of view of correlations suggesting that it should become a high-temperature superconductor if metallization is {achieved} by doping. We present also a computation of the Hubbard $U$ in density functional ``$+U$'' methods and discuss its relation to the Hubbard $U$ in spectroscopies. 
\end{abstract}

\maketitle

%\tableofcontents

\section{Introduction}

Since the discovery of high-$T_c$ superconductivity in copper oxides there has been intense efforts to find compounds with similar characteristics but different ions\cite{Norman2016}. Such materials can provide insights in understanding unconventional high-$T_c$ superconductivity and may lead to new applications. Iron pnictides and chalcogenides share similarities with cuprates in the phase diagram but have prominent multiorbital physics absent in cuprates. 
A  recent step forward\cite{Li2019} is the discovery of superconductivity in Sr-doped NdNiO$_2${,} which in the undoped phase has formally the same $d^9$ configuration as cuprates. However, differently from cuprates{,} parent phases seem to be metallic and non magnetic\footnote{It is worth to notice that a large superexchange has been measured in the  related overdoped trilayer compound La$_4$Ni$_3$O$_8$, J. Q. Lin {\it et al.}, Strong Superexchange in a $d_{9-\delta}$ Nickelate Revealed by Resonant Inelastic X-Ray Scattering, Phys. Rev. Lett. {\bf 126}, 087001 (2021). } and the splitting between Ni $d$ and O $p$ levels seems to be significantly larger\cite{Botana2020,Botana2021}. Also, the $T_c$ value for this material is quite low, in stark contrast with cuprates.

Another interesting nearest neighbor of Cu in the periodic table is Ag; however, it has long been known that the charge-transfer gap in AgO is formally negative\cite{Tjeng1990} and indeed AgO is not even magnetic. A positive charge-transfer energy can be recovered replacing the oxygen ion 
with the only element that can oxidize it, namely fluorine\cite{Grochala2001}.
\ch{AgF2} is a layered material with similar topology to that of cuprates but larger buckling of planes
which should depress the superexchange $J$.
Experiments show that $J$ reaches 70\% of a typical cuprate; moreover, flattening of the \ch{AgF2} sheets would blast the $J$ value way above those measured for cuprates\cite{Gawraczynski2019,Grzelak2020}.
% Furthermore \ch{AgF2} has a somewhat larger charge transfer gap and very similar $dd$ excitations\cite{Bachar2021}.
From the theory side, estimates of  hopping integrals between the metal and ligand  for 
\ch{AgF2} are very similar to known values in cuprates. \cite{Gawraczynski2019,Yang2014,Yang2015,Miller2020,Davydov2021}

A major open question to establish the similarities between the two families is the strength of the intra-orbital Coulomb repulsion in silver, $U_{4d}$. In general it is expected that the more diffusive character of $4d$ with respect to $3d$ orbitals will make correlations less important. It has been argued\cite{Gawraczynski2019} that this may be partially compensated by less efficient screening in  \ch{AgF2}, as \ch{F^-} is less polarizable than \ch{O^2-}.  Another important question is the magnitude of the charge-transfer energy $\Delta_{pd}$ as fluorides are often very ionic compunds while cuprates are quite covalent.
Recent optical and resonant inelastic x-ray scattering 
(RIXS) measurements show a larger fundamental gap and simultaneously a similar or even larger degree of covalence in \ch{AgF2} than in cuprates\cite{Bachar2021}. This apparently paradoxical result was explained in terms of a large inter-site Coulomb repulsion.

To further determine the relevant electronic parameters of \ch{AgF2}, 
here we present a high energy spectroscopy study combining valence band x-ray photoemission spectroscopy (XPS) of \ch{AgF2} ($d^9$)
and its filled-shell partner 
\ch{AgF} ($d^{10}$), 
Auger-Meitner spectroscopy of \ch{AgF} and cluster computations.

To determine the on-site repulsion directly from spectroscopic information one needs to access an excited state with two holes on the transition metal ($d^8$). This is naturally achieved in closed shell systems such as AgF by
Auger-Meitner spectroscopy\cite{Cini1977,Sawatzky1977}. Thus for this compound  we  obtained both $\Delta_{pd}$ and $U_{4d}$  using also valence band XPS.
While the \ch{AgF} $U_{4d}$ value can be taken as a reference{,} it would be desirable to have
spectroscopic information directly on \ch{AgF2}.  Valence band XPS does 
%prove
{probe}
the $d^8$ state of \ch{AgF2}. Unfortunately, present experiments do not resolve directly  $d^8$ satellites and the visible features do not allow {for} an accurate determination of parameters. 
%However, 
{Yet} we argue that the experiments impose enough constraints to classify the system as a strongly correlated charge-transfer insulator{,} confirming the analogy with cuprates.

%To address the above questions the spectra where compared with  computations in small clusters reflecting the local symmetry of the transition metal and using a DFT derived hybridizations\cite{Haverkort2012}.Cini-Sawatzky theory allows to determine the on-site repulsion of closed shell systems from   Auger-Meitner spectroscopy. 

 {\em Experimental Setup.} % Valence, core level photoemission spectroscopy and Auger-Meitner spectroscopy %(AMS)% 
 X-ray photoelectron spectra were measured using a custom-designed system made by SPECS (Berlin) with a loadlock connected to MBraun LABSTAR glovebox filled with Ar (monitored \ch{O2} and \ch{H2O} levels, typically $< 0.5$ ppm). 
 %An electron flood gun FG 15/40 was used for the charge compensation. % ; this device allows operation at the electron energies up to 500 eV and currents up to 1 mA. 
 The spectra presented in this paper were recorded with the Al-K$\alpha$ line (1486.74 eV, resolution 0.25 eV) at 32 mA emission current and 12.5 kV anode bias (400 W). A single crystal quartz mirror monochromator, the Phoibos 100 hemispherical analyser (100 mm mean radius) and a delayline electron detection system (DLD 3636) were used for the measurements. Powder samples were loaded in a holder provided by SPECS via a glove box in argon atmosphere. The \ch{AgF2} samples were sputtered with Ar+ ions at 6 mA and 2 keV for 30 min before spectra acquisition. 
An electron flood gun FG 15/40 was used for the charge compensation of the \ch{AgF2} samples.
More details can be found in Ref.~\onlinecite{Grzelak2015}.

\section{Model}
\begin{figure}[tb]
\centering
\includegraphics[width=8cm]{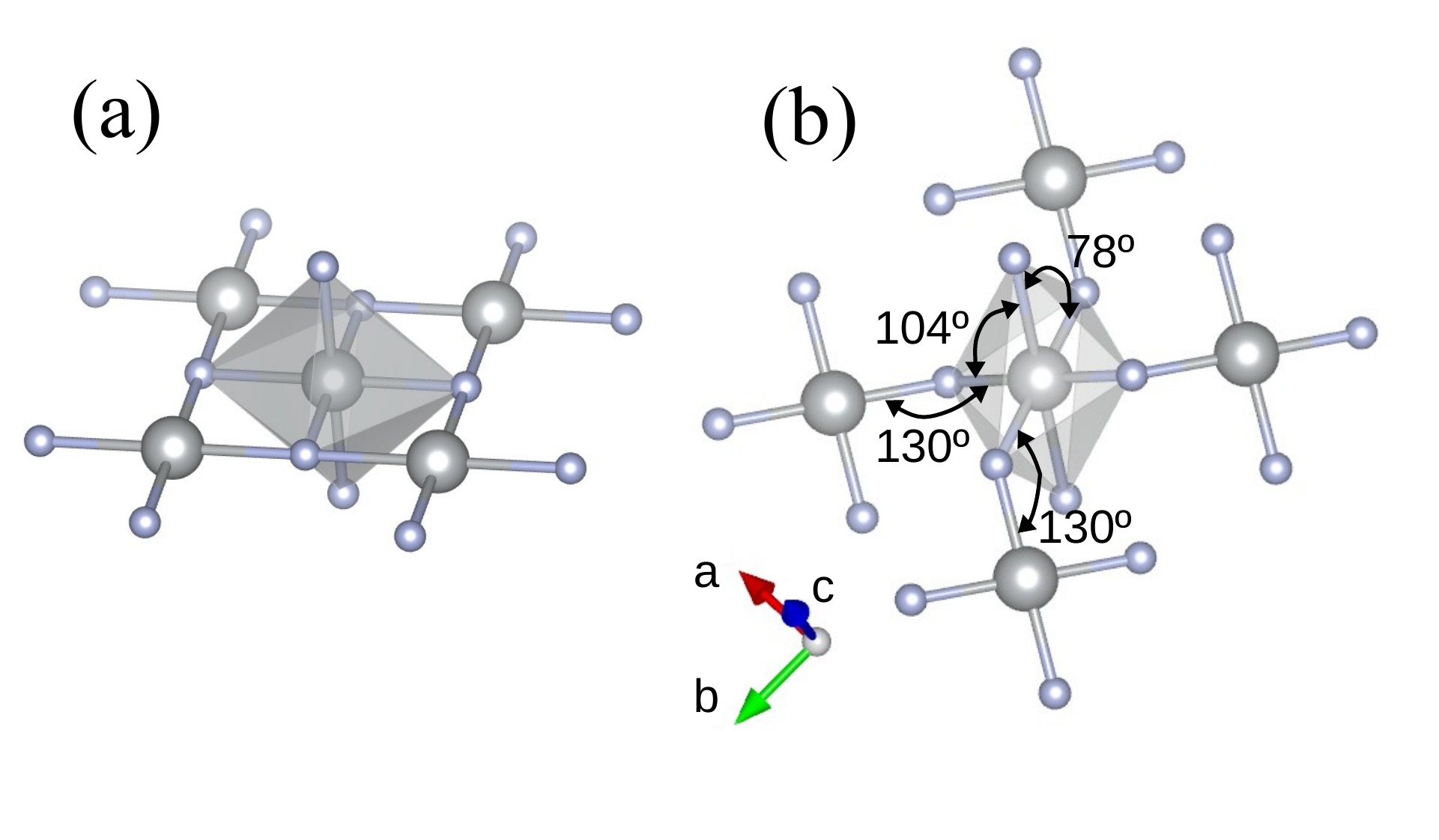}
\caption{\label{fig:octa}\ch{AgF6} clusters used in the computations (highlighted in gray) for AgF (a) and \ch{AgF2} (b). In the case of \ch{AgF2} we highlighted some angles. Analogous angles in AgF are either $90^\circ$ or $180^\circ$.  }
\end{figure}

For the cluster simulations we considered a central Ag atom 
and 6 surrounding F atoms.  Figure~\ref{fig:octa} shows 
the clusters used in gray and some extra surrounding atoms that are implicitly taken into account in the effective parameters defining the cluster Hamiltonian.
For AgF the local symmetry is $O_h$ which is well represented by an \ch{(AgF6)^{5-}} cluster [Fig.~\ref{fig:octa}(a)].
For \ch{AgF2} the point-group symmetry at the Ag site is $C_i$ (only inversion as non-trivial symmetry operation) thus we considered the distorted \ch{(AgF6)^{4-}} cluster shown in Fig.~\ref{fig:octa}(b).
We considered five $4d$ orbitals which hybridize with five 
% orbitals consisting of 
symmetry adapted linear combinations of $p$-orbitals from the neighboring F atoms as in Ref.~\onlinecite{Bachar2021}.

The cluster Hamiltonian in both cases reads, 
%We neglected small distortions and apical fluorines in \ch{AgF2} and took an \ch{(AgF4)^{2-}} cluster with $D_{4h}$ (square planar) symmetry. 
\begin{eqnarray}\label{eq:h}
 H &=& \sum_{\nu} \varepsilon_{4d}^\nu   d^\dagger_{\nu}  d_{\nu} +\sum_\nu \varepsilon_P^\nu  P^\dagger_\nu  P_\nu \nonumber\\
&+& \sum_{\nu} T_{pd}^\nu\left( {d}^\dagger_{\nu}  P_{\nu} +  P^\dagger_\nu  d_\nu \right)\nonumber \\
&+& \sum\limits_{\substack{\nu_1,\nu_2 \\ \nu_3,\nu_4}} U^{(PP)}(\nu_1,\nu_2,\nu_3,\nu_4)  {P}^\dagger_{\nu_1}  {P}^\dagger_{\nu_2} {P}_{\nu_3} {P}_{\nu_4}\\
   &+& \sum\limits_{\substack{\nu_1,\nu_2 \\ \nu_3,\nu_4}} U^{(dd)}(\nu_1,\nu_2,\nu_3,\nu_4)  {d}^\dagger_{\nu_1}  {d}^\dagger_{\nu_2} {d}_{\nu_3} {d}_{\nu_4}\nonumber\\
&+&\sum\limits_{\substack{\nu_1,\nu_2 }} U_{pd}  {P}^\dagger_{\nu_1} {P}_{\nu_1} {d}^\dagger_{\nu_2} {d}_{\nu_2}.  \nonumber
\end{eqnarray}
Here, $ d^\dagger_\nu$ creates a hole in the $4d$ orbitals while $ P^\dagger_\nu$ creates a hole in symmetry adapted combinations of F $p$-orbitals. $\nu$ labels orbitals and spin quantum numbers. $T_{pd}^\nu$ are hybridization matrix elements between the localized $d$-orbitals and the symmetry adapted $P$-orbitals.
% MNG Notation consistency: why not U_{PP} and U_{dd} ??

The procedure to define the symmetry adapted orbitals for  \ch{AgF2} {[Fig.~\ref{fig:octa}(b)]}
is explained in Ref.~\onlinecite{Bachar2021}. Despite the low local {$C_i$} 
symmetry, the hybridization matrix element{s} with the $P$-orbitals were found in density functional theory (DFT)\cite{Bachar2021} to be very close to matrix elements assuming $D_{4h}$ symmetry and a Slater-Koster\cite{Slater1954} parametrization with   
$T_{pd}^{x^2-y^2}=2.76$ eV{,} which is in good agreement with {typical} values in cuprates (2.3-3.0 eV).

The diagonal energies $\varepsilon_d^\nu$ and $\varepsilon_P^\nu$ from DFT are shown in Table~\ref{DFT_par_table}. In contrast with the hybridization {matrix elements}, 
$\varepsilon_P^\nu$ values are very different to the ones obtained in $D_{4h}$ symmetry. By construction the hybridization maximizes the overlap with the central Ag site which produces a quasi $D_{4h}$-symmetric set of parameters, but
the effect of the lower symmetry {is reflected} on diagonal energies. 
 Thus, for the latter symmetry constraints do not help to reduce the number of {independent} parameters. Furthermore, other problems preclude an accurate determination from first principles: First, 
$\varepsilon_P^\nu$ should also incorporate mean-field corrections to take into account  interaction with the next shell of atoms which is not present in the cluster. Second,  one expects that crystal field parameters will be more affected by self-interaction corrections than hybridization matrix elements. Indeed, crystal field splittings in DFT tend to be significantly smaller than in more accurate quantum chemistry methods\cite{Hozoi2011,Huang2011}.
  In view of these difficulties, we also explored an {\em ad hoc} set of diagonal energies as explained in Section \ref{sec:strongZR}. 

\begin{table}[t]
\caption{\label{DFT_par_table} 
  Crystal fields and hybridizations for \ch{AgF2} from the DFT computations of Ref.~\cite{Bachar2021}. 
  %MNG changed all \epsilon to \varepsilon in this Table
  We defined $\varepsilon_P^\nu=\Delta_{pd}+e_P^\nu$. All values are in eV.
The last three columns are the expression for a planar cluster  neglecting crystal fields in the $d$-shell from Ref.~\onlinecite{Eskes1990}}
\begin{ruledtabular}
\begin{tabular}{cccc|ccc}
       & &\ch{AgF2} ($C_i$)      &         & &\ \ \ \ $D_{4h}$& \\\hline
  $\nu$  & $\varepsilon_d^\nu$& $e_{P}^\nu$&$T_{pd}^\nu$& $\varepsilon_d^\nu$&$e_P^\nu$& $T_{pd}^\nu$ \\ \hline
 $z^2$   & -0.25          & 0.32   &  1.51 &0&$\frac45 T_{pp}$     &$\frac{1}{\sqrt{3}}T_{pd}^{x^2-y^2}$\\
 $x^2-y^2$& -0.28         & -0.16  &  2.76 &0&$-\frac65 T_{pp}$ &${T_{pd}^{x^2-y^2}}$  \\
 $x y$   & 0.34           &  -0.05 &  1.36 &0&$\frac45 T_{pp}$ &$\frac{1}{2}{T_{pd}^{x^2-y^2}}$\\
 $x z$   & 0.09           &  -0.14 &  1.05 &0&$-\frac15 T_{pp}$ &$\frac{1}{2 \sqrt{2}}{T_{pd}^{x^2-y^2}}$\\
 $y z$   & 0.10           &   0.04 &  1.02  &0&$-\frac15 T_{pp}$ & $\frac{1}{2 \sqrt{2}}{T_{pd}^{x^2-y^2}}$ \\
 \end{tabular}
\end{ruledtabular}
%\vspace{1ex}
%     {\raggedright $^\dagger$ ($x^2-y^2$) symmetry.  \par}
\end{table}

%The one particle Hamiltonian for \ch{AgF2} was obtained from DFT computations  followed by projection onto maximally localized Wannier orbitals as explained in Ref.~\onlinecite{Bachar2021}. 
%The Wannier basis was then truncated to the cluster and non-bonding $p$ orbitals were eliminated to reach a basis of 5 combination of $p$ orbitals with approximately the same symmetry of the $d$ cubic harmonics on Ag. 
%Local axis on the Ag site are approximately in the direction of the planar Ag-F bonds and the orbitals have, with a good approximation, the symmetry of cubic harmonics transforming as  ${3z^2-r^2}$, ${x^2-y^2}$, ${xy}$, ${xz}$ and ${yz}$.

%Once $T_{pd}^{x^2-y^2}$ is fixed in Table~\ref{DFT_par_table}, the other values are in good agreement with the expressions used in Ref.~\onlinecite{Eskes1990} for a $D_{4h}$ planar configuration. Also the absolute value of  $T_{pd}^{x^2-y^2}=2.76$eV is in good agreement with values in cuprates (2.3-3.0 eV).

%On the other hand, the splittings  $e_P^\nu$ are not in agreement  with the  $D_{4h}$ expressions showing that the lowering of the symmetry to $C_i$ affects mainly the $P$ shell.   

For AgF,  the definition of the symmetry adapted orbitals is standard.
The F-Ag hopping matrix element, $t_{pd}=T_{pd}^{x^2-y^2}/2$ has been shown to follow Andersen scaling\cite{Andersen1978}
$t_{pd}'=t_{pd}(d/d')^4$ and $t_{pp}=t_{pp}(d/d')^2$ with $d$ the distance between the involved atoms (see Supplementary Information to Ref.~\onlinecite{Gawraczynski2019}). Thus for AgF we adapted the hybridization matrix elements derived for \ch{AgF2} with the Andersen scaling correction yielding  $T_{pd}^{x^2-y^2}= 1.38$ eV and $T_{pp}= 0.29$ eV.

With these settings a total of 20 electrons can be accommodated in the clusters for both fluorides. Stoichiometric AgF and \ch{AgF2} correspond to configurations with zero ($N=20$ electrons) 
and one hole respectively ($N=19$). 
We define the charge-transfer energy as the energy cost $\Delta_{pd}=E(d^{10}\underline{L})-E(d^{9})$ for $T_{pd}^\nu=0$ with  
$\underline L$ denoting a hole in the ligand. {The energies $E$} are averages of the corresponding multiplets including the diagonal energies $\varepsilon_d^\nu$ and $\varepsilon_P^\nu$.

%Following Ref.~\onlinecite{Ghijsen1988} we neglect crystal fields respect to hybridization effects, thus $\varepsilon_{4d}$ is independent of $\nu$ and   
% $\varepsilon_P(\nu)$ is split by the $pp$-hybridization. % which split the symmetry adapted orbitals and includes the Coulomb repulsion in F in mean-field.
%  $T_{pd}(\nu)$ is the $d-P$ transfer integral and can be expressed in terms of Slater-Koster\cite{Slater1954} two-center integrals ($pd\sigma$) and ($pd\pi$). 

When referring to the Hubbard $U$-repulsion it is important to specify which Hubbard {interaction} is meant.  
In the two-hole subspace $U^{(dd)}$ is a $45\times45$ matrix representing Coulomb repulsion on Ag. From the trace we can define the average interaction $\overline U_{dd}=F^0-2(F^2+F^4)/63${,} with $F^k$ {being the} Slater-Condon integrals. Another important matrix elements is the intra-orbital repulsion $U_{4d}\equiv U(^1A_{1g})=F^0+4 (F^2+F^4)/49$ where the symetry in parenthesis correspond to the $d^8$ state. The Auger process is dominated by a  $^1G$ final state\cite{Tjeng1990} so it is useful to define also 
$U(^1G)=F^0+(36F^2+F^4)/441$.

$F^2$, $F^4$ represent high multipole Coulomb interaction matrix elements and are {weakly} screened in solids. They can be obtained from other compounds with \ch{Ag^2+} ions. We took $F^2=8.19$~eV and $F^4=6.80$~eV from Ref.~\onlinecite{Tjeng1990}.
On the other hand, $F^0$ is strongly screened in an environmental dependent manner\cite{Antonides1977,DeBoer1984,Meinders1995,Brink1995} and our goal is to determine it{,} or {equivalently} $U_{4d}${,} in silver fluorides. The latter is a good estimate 
of the  repulsion parameter  in a generalized Emery model\cite{Emery1987} to be used as a starting point of the theoretical description of correlation in  \ch{AgF2}. 
Because of the $d^9$ nature of \ch{AgF2}{,} one expects that this model 
captures both the intermediate-energy {physics} (excitons, optics, magnetic interactions) {as well as} the low-energy physics. 

Onsite F and inter-site Ag-F Coulomb repulsions can also be large in this system as F orbitals are more localized than O {orbitals}, 
so they were explicitly retained instead of absorbing them {at a mean-field level} as {it is} often done in cuprates.
The repulsion in a single fluorine is described by an $U^{(pp)}$ matrix. For simplicity we retained only one parameter, the monopole {term} of $U^{(pp)}$.
This interaction was projected on the basis of symmetry adapted $P$ orbitals which resulted in the $U^{PP}$ matrix as described in Appendix~\ref{sec:treatm-fluor-repuls}. 
For $U_{pd}$, for simplicity, we also retained only the monopole {term} which leads straightforwardly to the expression in Eq~(\ref{eq:h}).

%HOW CAN WE PARAMETRIZE THIS AND HOW IT IS RELATED TO $U_{pp}$ ON-SITE?

%For the interaction part we went beyond Ref.~\onlinecite{Bachar2021} and considered intershell F-Ag ($U{Pd}$) and intrashell F ($U_{pp}$) interactions. 

%Past experience in oxides\cite{Haverkort2012} have shown that DFT computations provides accurate values for hybridization integrals therefore to minimize the unknowns parameters we fix hybridizations to previous estimates\cite{Gawraczynski2019}, namely $t_{pp}=0.3$eV for the nearest neighbor F $p$ hopping and
%$t_{pd}= 1.38$ eV for the hopping between a $d_{x^2-y^2}$ orbital and a  $p$ orbital pointing to one $d$-orbital lob in \ch{AgF2}.  

%Taking  for the matrix elements between Ag and F centered orbitals for AgF we took approximate hybridization values using Andersen scaling\cite{Andersen}   This was checked to be more accurate than Harrison\cite{Harrison1989} scaling in the SI of Ref.~\onlinecite{Gawraczynski2019}.
%The relation of these parameter with the ones in Eq.~(\ref{diagonal}) is %standard and is given in the Supplementary Information. 

The valence band photoemission spectroscopy of AgF and \ch{AgF2} was computed in terms of the one hole spectral function of the cluster projected on $d$ or $P$ states,  $A_{d/P}(\omega)$. This should be weighted with the atomic cross section ratios which were taken from Ref.~\onlinecite{Yeh1985} (divided by the number of electrons in the shell). In the case of the Al-K$\alpha$ source used in the present study, one obtains $r=\sigma(\ch{F} 2p)/\sigma(\ch{Ag} 4d)=0.05$.

Because the number of F per Ag is different in the cluster and the bulk compound {an additional} correction is needed.  
In the ionic limit, the total integrated F $p$-spectral weight per Ag atom corresponds to 12 states per \ch{AgF2} and 6 states for AgF. Instead, in the clusters those weights are 10 in both cases corresponding to 10 filled F orbitals.  To take this into account, we computed the photoemission intensity as $I=A_{d}(\omega)+r (12/10) A_{P}(\omega)$ for \ch{AgF2} and 
$I=A_{d}(\omega)+r (6/10) A_{P}(\omega)$ for \ch{AgF}. In practice due to the smallness of $r$ the intensity is dominated by the silver response. 
Following Ref.~\onlinecite{Ghijsen1988} we assumed that the Auger-Meitner spectra is dominated by the two-hole spectral function in the singlet channel\cite{Tjeng1990}.

\begin{figure}[tb]
\centering
\includegraphics[scale=0.55]{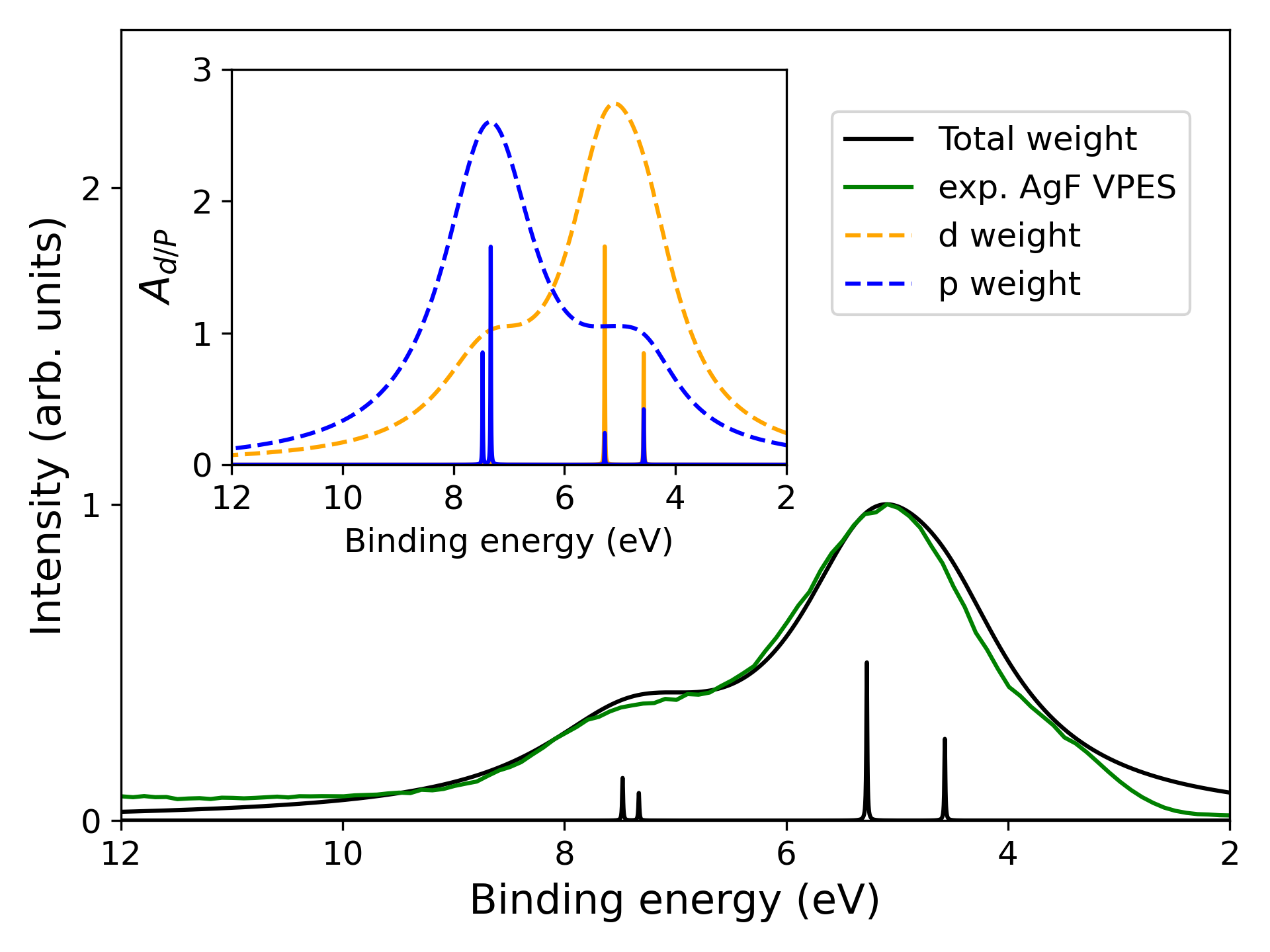}
\caption{\label{AgF_VPES} Comparison between the valence XPS experimental spectrum (green) and cluster model calculation (black) of the \ch{AgF} compound.
  In the computations we fixed hybridization matrix elements as explained in the text
yielding $T_{pd}^{x^2-y^2}= 1.38$ eV and $T_{pp}= 0.29$~eV.  
The best fit to the experiment was found with $\Delta_{pd}= 1.2$~eV.  We show the computation with a very narrow broadening to resolve the position of individual excitations and intensities and with a phenomenological Lorentzian broadening growing linear in energy ($\eta = 0.97$~eV $ + 0.14 \omega $)  to fit of the experiment. The inset shows the computed silver-$d$ (orange) and fluorine-$P$ (blue) spectral functions.
}
\end{figure}

% \begin{table}[h]
%   \caption{\label{param} The \ch{AgF} parameters in eV used in the simulation of the \ch{(AgF6)^{4-}} cluster. The Slater-Condon integrals correspond to the Racah parameters of Ref.~\onlinecite{Tjeng1990}}
% \begin{ruledtabular}
% %\begin{tabular}{cccccc}  {$\Delta_{pd}$}&{$T_{pd}(e_{g})$}&{$T_{pp}$}&{$\overline U_{dd}$}&{$B$}&{$C$} \\
% \begin{tabular}{cccccc}  {$\Delta_{pd}$}&{$T_{pd}(e_{g})$}&{$T_{pp}$}&{$\overline U_{dd}$}&{$B$}&{$C$} \\
%   \hline \\
% 1.2 & 1.38 & 0.29 & 6.0 & 0.09 & 0.54 
% \end{tabular}
% \end{ruledtabular}
% \end{table}

\section{ Results}

\subsection{Valence band photoemission of AgF}
We start  discussing the valence band photoemission spectra of AgF, which is shown in Fig.~\ref{AgF_VPES}. The experimental spectrum is close to previous published results\cite{Grochala2003}. Since AgF is a filled shell system,   the final state has only one hole which makes two-hole interactions irrelevant.
In this case, for simplicity we follow Ref.~\onlinecite{Eskes1990} and fix
$\varepsilon_{4d}^\nu=0$ and   $\varepsilon_P^\nu= \Delta_{pd}$. Using the hybridization determined above and
varying $\Delta_{pd}$ we obtain excellent agreement with the experiment.

The inset shows the removal spectral functions contributing to the line shape as explained above. The $d$($p$)-spectral weight peaks near 5 (7.5) eV binding energy and shows strong covalence\cite{Grochala2003}. Indeed, 
there is a sizable shoulder of the $d$ spectral function at the peak of the $p$ spectral function and vice versa. This is reflected also in the small value of the charge-transfer energy for this compound, $\Delta_{pd}=1.2$ eV,
and illustrates nicely that silver fluorides are exceptionally covalent~\cite{Grochala2003} in contrast to most fluorides which are ionic. 

\begin{figure}[t]
\centering
\includegraphics[scale=0.55]{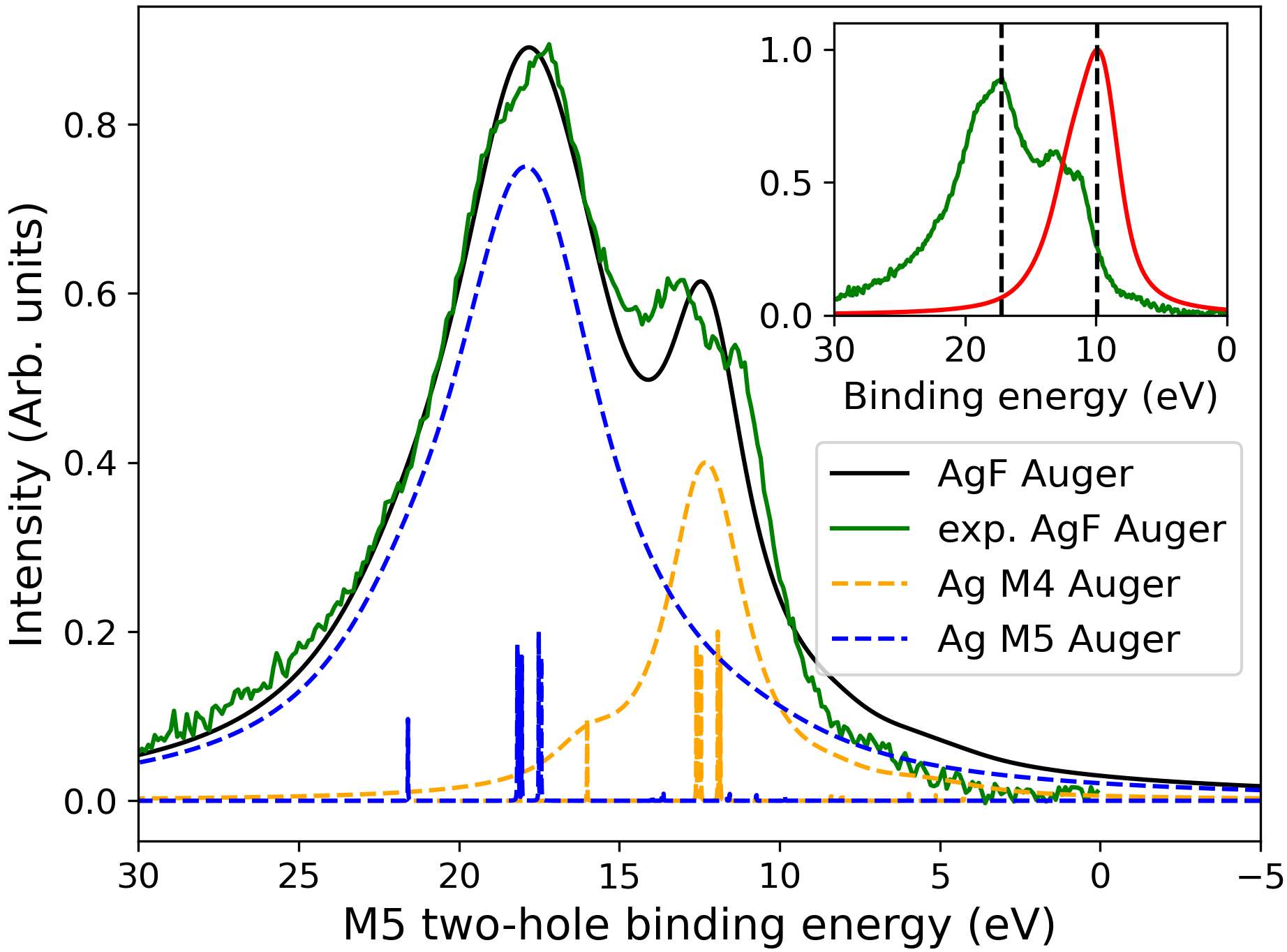}
\caption{Comparison between Auger experimental spectrum (green) and Auger cluster model calculation spectrum (black) of the \ch{AgF} compound. The black curve is given by the summation of the $M4$ and $M5$ singlet spectra. The two main structures are due to the SO splitting in the Ag $3d$ subshell (see Appendix.~\ref{app:core-hole-spectra}).
  The broadening of the $M4$ component is $\eta = 2.6$ eV and the one of the $M5$ component is $\eta = 5.6$ eV. The fit between experimental and theoretical data fixes $\overline U_{dd}=6.0$ eV. Other cluster parameters are the same as in Fig.~\ref{AgF_VPES}.
  In the inset, the self convolution of the experimental valence band XPS spectrum (red) and the experimental Auger spectrum (green) are shown. The energy splitting (dashed black vertical lines) between those two peaks is equal to 7.3 eV and represents a rough estimate of $U(^1G)$ in the \ch{Ag} $4d$ subshell.  
}
\label{Auger_tot}
\end{figure}

\subsection{Auger-Meitner spectroscopy of AgF}
The above experiment sets the stage for Auger-Meitner
$MNN$ spectroscopy (hereafter Auger spectroscopy) in AgF
which is well suited to probe two-hole interactions in silver\cite{Cini1976,Sawatzky1977} in a fluorine environment, just as \ch{Ag2O} and \ch{Cu2O} where used for the same purpose in an oxygen environment\cite{Tjeng1990,Ghijsen1988}. 

The green line in Fig.~\ref{Auger_tot} shows the $MNN$  Auger spectra of AgF which is better resolved than previous measurements\cite{Wolan1998}.
In this process, a core hole is created in the $3d$ shell which is filled by a $4d$ electron with the simultaneous ejection of a second $4d$ electron resulting in a formally $d^{8}$ final state,
 whose energy is determined by the hole-hole Coulomb repulsion.
Indeed, if the two-holes were not correlated one should find a spectral function given by the convolution of the one-hole spectral function of Fig.~\ref{AgF_VPES}. Instead, for correlated holes one finds a significant shift\cite{Powell1973,Sawatzky1977,Cini1976,Seibold2008} which is approximately given by $U(^1G)$ on silver. This is shown in the inset of Fig.~\ref{Auger_tot} which compares the convoluted one-particle spectra (red line) and the Auger spectra with the binding energy computed assuming an $M5$ initial state (see Appendix \ref{app:core-hole-spectra}). 
 The distance between the peaks is 7.3 eV{,} which provides a first estimate of $U(^1G)$.

\begin{figure}[tb]
\centering
\includegraphics[scale=0.55]{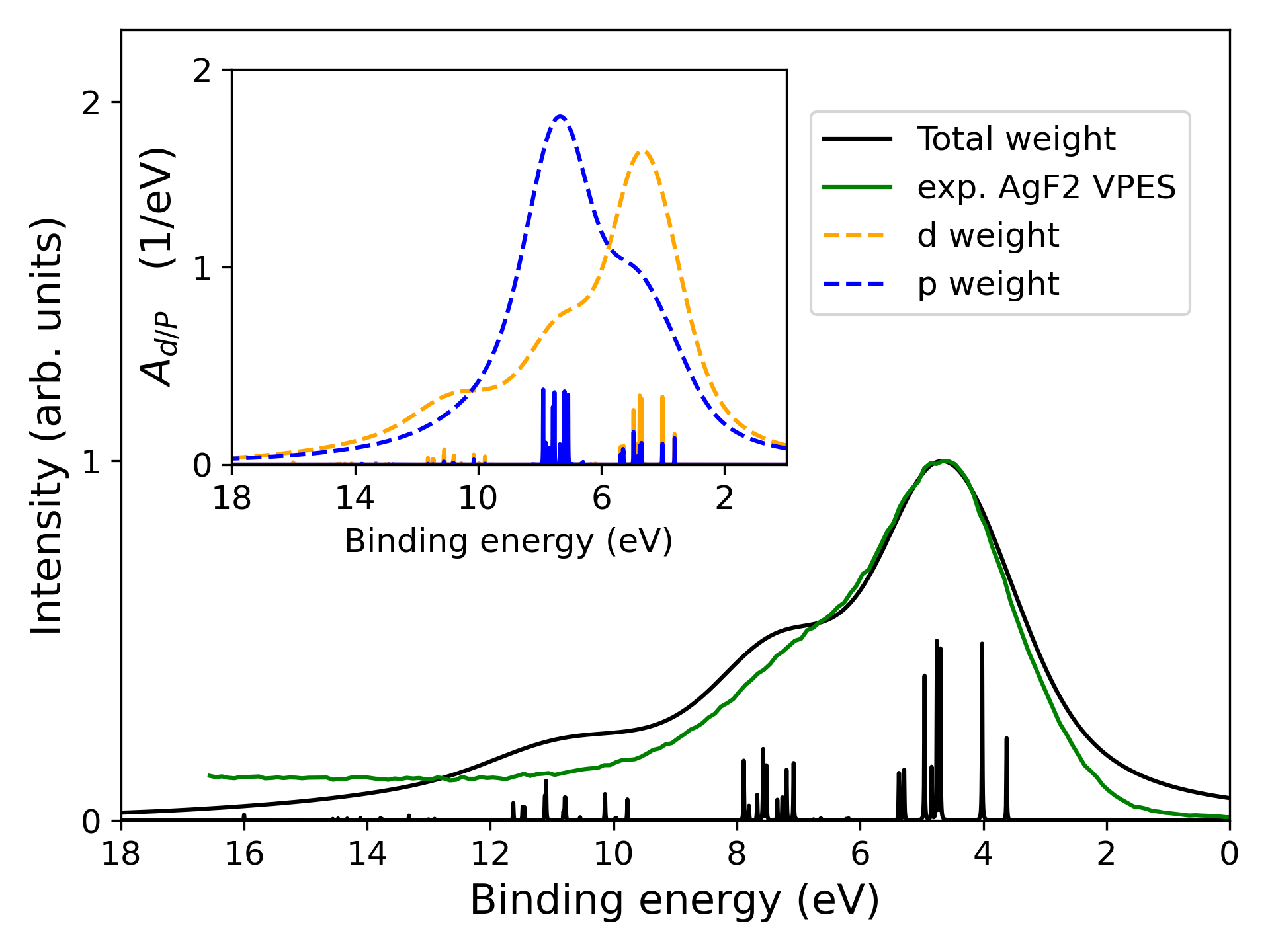}
\caption{\label{fig:agf2_vpes_dft}
  \ch{AgF2}
  Valence band XPS spectra of AgF$_2$. The green line is the experimental result and the black line is the result of the cluster computation with the crystal field parameters of Table~\ref{DFT_par_table} and  
  $\overline U_{dd}=5.5$ eV ($U_{4d}=7.2$eV) , $\overline U_{pp} = 6$ eV, $U_{pd} = 1$ eV, $\Delta_{pd} = 4$ eV, $T_{pd} = 2.76$ eV. We used a Lorentzian broadening $\eta = 0.83 + 0.17\,\omega$ eV and  convolution with a Gaussian of FWHM = 1.0 eV (green line). We also show the theory with a very small broadening to show the energy of individual states in the cluster (black line). The inset shows the computed character of the spectra: $A_d(\omega)$ (orange dashed line) and $A_P(\omega)$ (blue dashed line). 
For the theory, the zero of the energy is set at the  addition chemical potential. 
For the experiment, the spectra was shifted in energy to align its main features to the cluster computation.
}
\end{figure}

A more accurate value of the silver onsite repulsion can be obtained with the cluster computation which is show{n} in the main panel {of Fig.~\ref{Auger_tot}} and compared with the experimental result. From the fit we find $\overline U_{dd}=6.0$ eV. This corresponds to $U(^1G)=7.16$~eV which is close to the above estimate and can be compared with $U(^1G)=5.8$ eV for the closed-shell {\em oxide} \ch{Ag2O} found by  
Tjeng, Sawatzky and collaborators\cite{Tjeng1990} using the same technique. The intra-orbital repulsion in AgF amounts to $U_{4d}=7.7$~eV.

In both silver compounds the Hubbard repulsion is much smaller than for a free  \ch{Ag^2+} ion [$U(^1G)=14.8$~eV] due to the strong screening by the environment\cite{DeBoer1984,Marel1988,Meinders1995,Brink1995,Sawatzky2009}. Furthermore, 
the different on-site repulsion between AgF and  \ch{Ag2O} can be attributed to differences in that  screening.
Indeed, \ch{F^-} is roughly a factor of 2 less polarizable~\cite{DeBoer1984} than \ch{O^2-} and the Ag-F distance is 2.468~\AA, larger than  {the Ag-O bond length of 2.05~\AA}. Since the decrease in the on-site interaction is proportional to the polarizability of the environment and decreases with the metal-ligand distance to the fourth power, we conclude that both effects lead to less screening (larger repulsion) on the fluoride. On the other hand,  \ch{Ag2O} is linearly coordinated with 2 \ch{O^2-} while \ch{AgF} is octahedrally coordinated with 6 \ch{F^-}{,} which may partially compensate this effect. Clearly{,} the {reduced} screening prevails so the  $U(^1G)$ value is significantly larger in AgF {than in} \ch{Ag2O}. This is smaller than $U(^1G)=9.2$~eV{, the value} found~\cite{Ghijsen1988} in closed shell \ch{Cu2O} but {it is } still sizable. 

%square planar coordinated with 4 \ch{F^-} with metal ligand-distances roughly the same as in AgF. This geometry is expected to lead to higher screening in \ch{AgF2}.

%The two-hole final state has dominant singlet character ($^1G$) with a $\ket{d_{x^2-y^2} d_{xy}}$ orbital wave function giving access to $U_{4d}(^1G)=U_{4d}(A_{1g})-C$ with $U_{4d}(A_{1g})=A + 4 B + 3 C$ the intraorbital Hubbard repulsion and  $A,B,C$ are Racah parameters ($C=0.54$ eV for Ag)\cite{Tjeng1990}. 

%A fraction of that screening is attributed to the surrounding ions\cite{Meinders1995,Brink1995} but \ch{F^-} is roughly a factor of 3 less polarizable than \ch{O^2-} which may lead to larger effective repulsion. 
% A lower bound for $U_{4d}$ is given by analysis of Auger MNN data for elementary Ag in metallic form\cite{Powell1973}. A core hole is created in the $3d$ states which is filled by a $4d$ electron with the simultaneous emission of a second $4d$ resulting in a two-hole final state. 

%Because of the large spin-orbit coupling the $3d$ core hole and consequently the Auger line shape  splits into levels M5 and M4 (energy separation, $\Delta_{SO}=6.0$ eV\cite{Grzelak2015}).
%We compute all binding energies assuming the initial hole is M5 so the M4 feature appears shifted to lower binding energies by $\Delta_{SO}$.

\subsection{Valence band Photoemission of \ch{AgF2} }

We now turn to the valence band photoemission spectra of \ch{AgF2}.
For an $N$-electron system with ground-state energy $E_0(N)$, 
photoemission {probes} %proves 
the many-body states with $N-1$ electrons which in our case is the Hilbert space spanned by the  $d^8$, $d^{9}\underline L$ and  $d^{10}\underline L^2$ configurations. We will discus two possible theoretical scenarios which are plausible at present: one using the DFT crystal fields %and 
yielding a weakly bound Zhang-Rice (ZR) state\cite{Eskes1988,Zhang1988}
%MNG add ZR citation
and another using an {\em ad hoc} set of crystal fields %and 
yielding a strongly bound ZR state.

\begin{figure}[tb]
\includegraphics[scale=0.5]{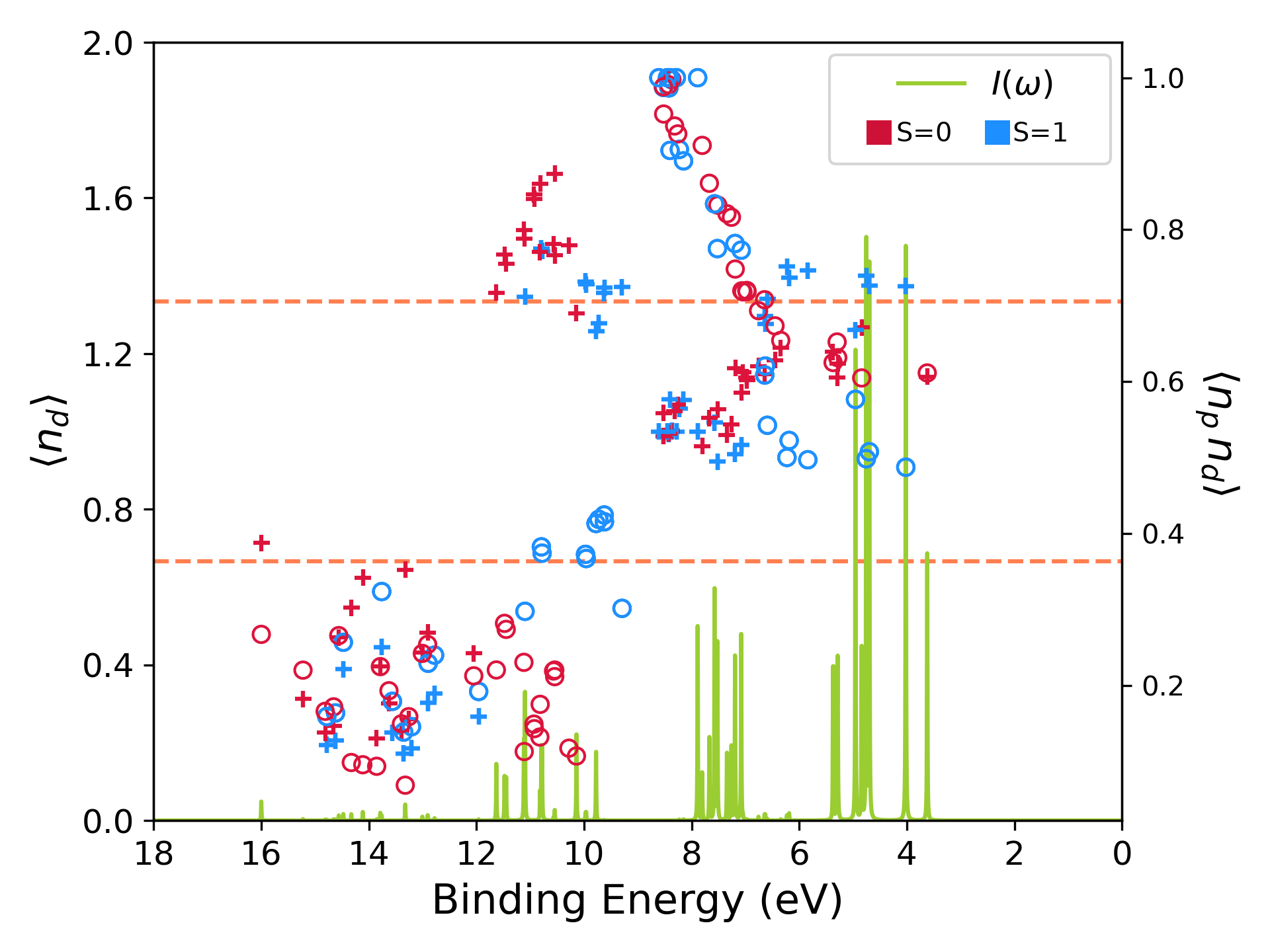}% Here is how to import EPS art
\caption{\label{fig:nddw_cf_dft}
The crosses are the hole occupancy  
  of the $d$-states (left scale) while the circles are the interatomic charge correlator (right scale) in the possible final states of the photoemission process. The color of the symbols encode the singlet (red) or triplet (blue) character of the two-hole state. 
  Parameters and binding energies are defined as in Fig.~\ref{fig:agf2_vpes_dft}. At the bottom we show the valence band spectrum on green to facilitate identification of the states.
  The horizontal lines are guide to the eyes separating large, small, and intermediate values of the expectation values.  
}
\end{figure}

\subsubsection{Weakly bound Zhang-Rice state scenario}\label{sec:weakZR}
The experimental spectra is shown in green in Fig.~\ref{fig:agf2_vpes_dft}. The black line is the cluster computation with the DFT crystal field energies from Table~\ref{DFT_par_table} and parameters indicat{ed} in the caption. In the inset we show the $d$- and $p$-spectral functions.
The theory  shows a main peak and a 
%shoulder somehow more intense
{somewhat more intense shoulder}
than in the experiment and a secondary shoulder which is {either} not visible in the experiment or %it is 
hidden by the background.

As will be shown below, the narrow theoretical peak (black line) with smaller binding energy (3.6 eV)
corresponds to the ZR state\cite{Eskes1988,Zhang1988}.  This is the first ionization state of the system, therefore its binding energy
coincides with the chemical potential to remove one electron defined as 
$\mu^-\equiv E_0(N-1)-E_0(N)$.  We {have used} the freedom to choose the zero of
the one particle energy levels in our computations in such a way that the addition chemical potential $\mu^+\equiv E_0(N+1)-E_0(N)$ coincides with zero binding energy. In this way, the ZR binding  energy yields also the charge gap of the system, $E_{gap}\equiv\mu^+-\mu^-=3.6$~eV in Fig.~\ref{fig:agf2_vpes_dft}.

In order to characterize the excited states, 
figure \ref{fig:nddw_cf_dft} shows $\langle n_d \rangle$, the hole occupancy in $d$ orbitals,
and the correlator $\langle n_d n_p \rangle$ of all possible two hole states in the cluster, as a function of their binding energy in the photoemission spectra. Here we defined,  
$n_d\equiv\sum_\nu {d}^\dagger_{\nu} {d}_{\nu}$ and $n_p\equiv \sum_\nu   {P}^\dagger_{\nu} {P}_{\nu} $. $n_d$ projects on the subspace spanned by $d^8$ and $d^9\underline L$ configurations while $ n_d n_p $ projects on the $d^9\underline L$ subspace. Clearly, these two expectation values take integer values in these subspaces: 
\begin{eqnarray*}
  &\braket{n_d}=2,& \quad \braket{n_d n_p}=0, \quad (d^8), \\
  & \braket{n_d}=1,&\quad \braket{n_d n_p}=1, \quad (d^9\underline L),\\
  &\braket{n_d}=0,&\quad \braket{n_d n_p}=0, \quad (d^{10}\underline L^2).  
\end{eqnarray*}
Therefore, for general states the expectation values and completeness 
characterize the probability of the three possible configurations.
For example a state with  $\braket{n_d}\approx 1$ may indicate a $d^9\underline L$ state or an equally mixed  $d^8$ and  $d^{10}\underline L^2$ configuration. 
The $ \braket{n_d n_p} $ correlator allows to distinguish among these two possibilities or intermediate situations. %Not clear: of --> from?
We use red and blue {in Fig.~\ref{fig:nddw_cf_dft}} to distinguish singlet and triplet final states respectively.  We also show the spectra as narrow peaks as a reference.

\begin{figure}[tb]
  \includegraphics[scale=0.55]{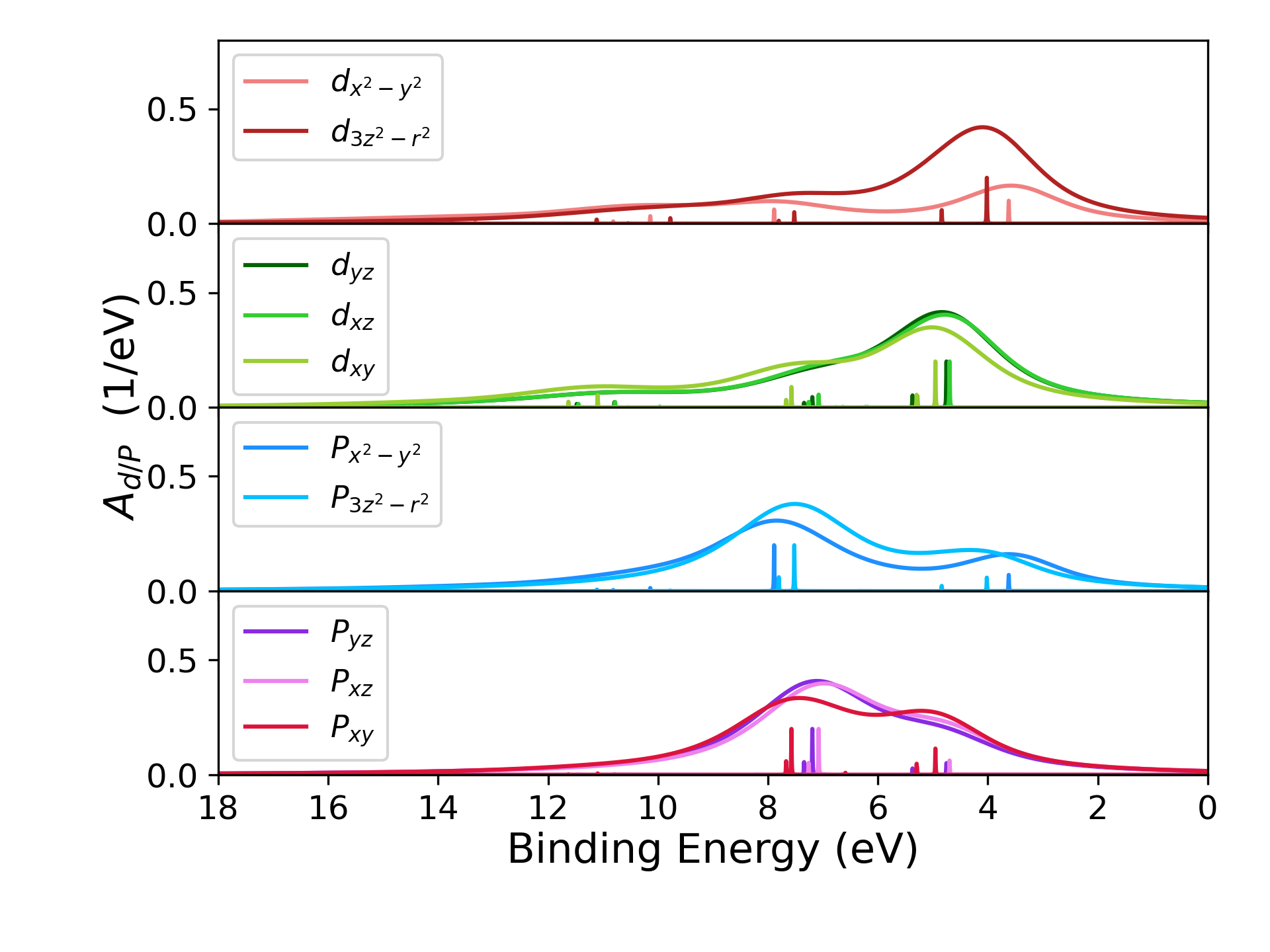}% Here is how to import EPS art
  \caption{\label{fig:Adpdw_cf_dft} One particle spectral function projected on the different orbitals symmetries of the cluster for the same parameters as Fig.~\ref{fig:agf2_vpes_dft}.    }
\end{figure}

The main peak in the {XPS} experiment {[see Fig.~\ref{fig:agf2_vpes_dft}]}
can be attributed to states with $\langle n_d \rangle\approx 1$ and also intermediate interatomic charge correlations indicating a mixed character (states between 3 and 6 eV).
The main shoulder (states between 6 and 8 eV)  is characterized by final states with large interatomic charge correlations indicating a prevalent  $ d^{9}\underline L$ character.
The secondary shoulder (states with energy between 9 and 12 eV) is due to states with large $\langle n_d \rangle$ and small interatomic charge correlations which indicates a prevalent $ d^{8}$ character.
Above 12 eV states with small $d$-spectral weight appear{, whose} %which have
small  $\langle n_d \rangle$  indicat{es} %ing
a $ d^{10}\underline L^2$ character. 

\begin{figure}[tb]
  \includegraphics[scale=0.5]{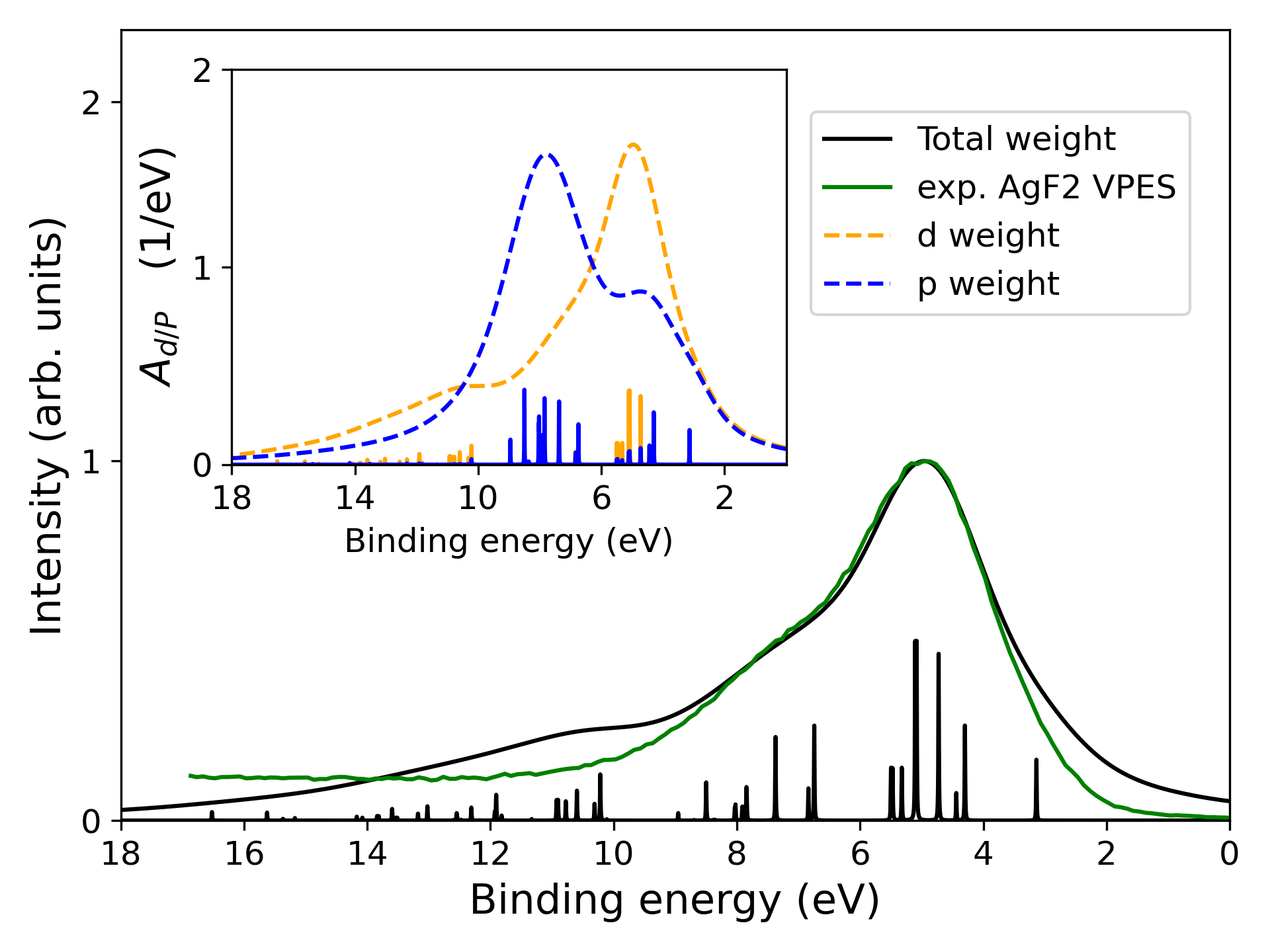}% Here is how to import EPS art
\caption{\label{fig:agf2_vpes} Valence band XPS spectra of AgF$_2$. The green line is the experimental result and the black line is the result of the cluster computation with $\overline U_{dd}=6.5$ eV corresponding to $U_{4d}=8.2$ eV. 
  Other parameters are  $\overline U_{pp}=6.0$ eV, $\Delta_{pd}=3.5$ eV,  $U_{pd}=1.5$ eV. 
For the theory, the zero of the energy is set at the  addition chemical potential. 
For the experiment, the spectra was shifted in energy to align its main features to the cluster computation.  The theoretical curve is presented with a phenomenological broadening with a Lorentzian parameter varying linearly with energy as $\eta = 0.45 {\rm eV} + 0.30 \,\omega$ (green line) and with a very small broadening to show the energy of individual states in the cluster (black line). The inset shows the computed character of the spectra: $A_d(\omega)$ (orange dashed line) and $A_P(\omega)$ (blue dashed line).}
\end{figure}

Figure \ref{fig:Adpdw_cf_dft} shows the orbital resolved spectral function. Combining with Fig.~\ref{fig:nddw_cf_dft} we see that the leading narrow peak  corresponds to a  singlet final state with orbitals 
 with dominant  $x^2-y^2$ character and  a sizable weight in the
$ d^{9}\underline L$ configuration. As anticipated{,} we can identify this state with the ZR singlet of cuprates.\cite{Eskes1988,Zhang1988}

One can see that the energy gap of the singlet state {with} respect to triplet, i.e. the difference of their binding energies, is quite small indicating that it competes with a
% prevalently
{predominantly} $d^8$ high-spin state\footnote{Some times this energy gap is defined as the ``ZR singlet binding energy''. Here we refrain to use this terminology as we use ``binding energy'' with a different meaning.} with the second hole in the $d_{z^2}$ orbital.  
The present parameter set has larger $\Delta$ and smaller $U_{4d}$ with respect to the parameter set to be analyzed in what follows (and {with respect to} cuprates). 
The small stability of the ZR singlet is in accord with the results of Ref.~\onlinecite{Eskes1988} where the stability of the singlet decreases with  $\Delta-U_{4d}$. In particular the present weakly bound ZR scenario is similar to a recent proposal for nickelates.\cite{Jiang2020} Note that the 
 $d^8$ high-spin state is the Hund-rule ground state in a scenario where the charge gap is of the Mott type in the Zannen Sawatzky Allen (ZSA) classification scheme.\cite{Eskes1988}
 Thus the present scenario is of the charge transfer type but close to the Mott regime.

% Unfortunately since the dominantly $ d^{8}$ features are not seen one can not obtain a direct measure of the Hubbard interaction.

 In general the relative weight of the shoulder {with} respect to the main peak
 decreases increasing  $\overline U_{pp}${,} thus the fitting improves for 
 $\overline U_{pp}=10$~eV{. H}owever the effect is rather small as $d^{10}L^2$ configurations do not appear directly in the spectrum{; as a consequence} this method lacks sensitivity to determine  $\overline U_{pp}$.  One can obtain a similar result increasing $\Delta_{pd}$ beyond 4 eV. These values, however would be in strong contradiction {to} the RIXS and optical experiments of Ref.~\onlinecite{Bachar2021}.

\begin{figure}[tb]
\includegraphics[scale=0.5]{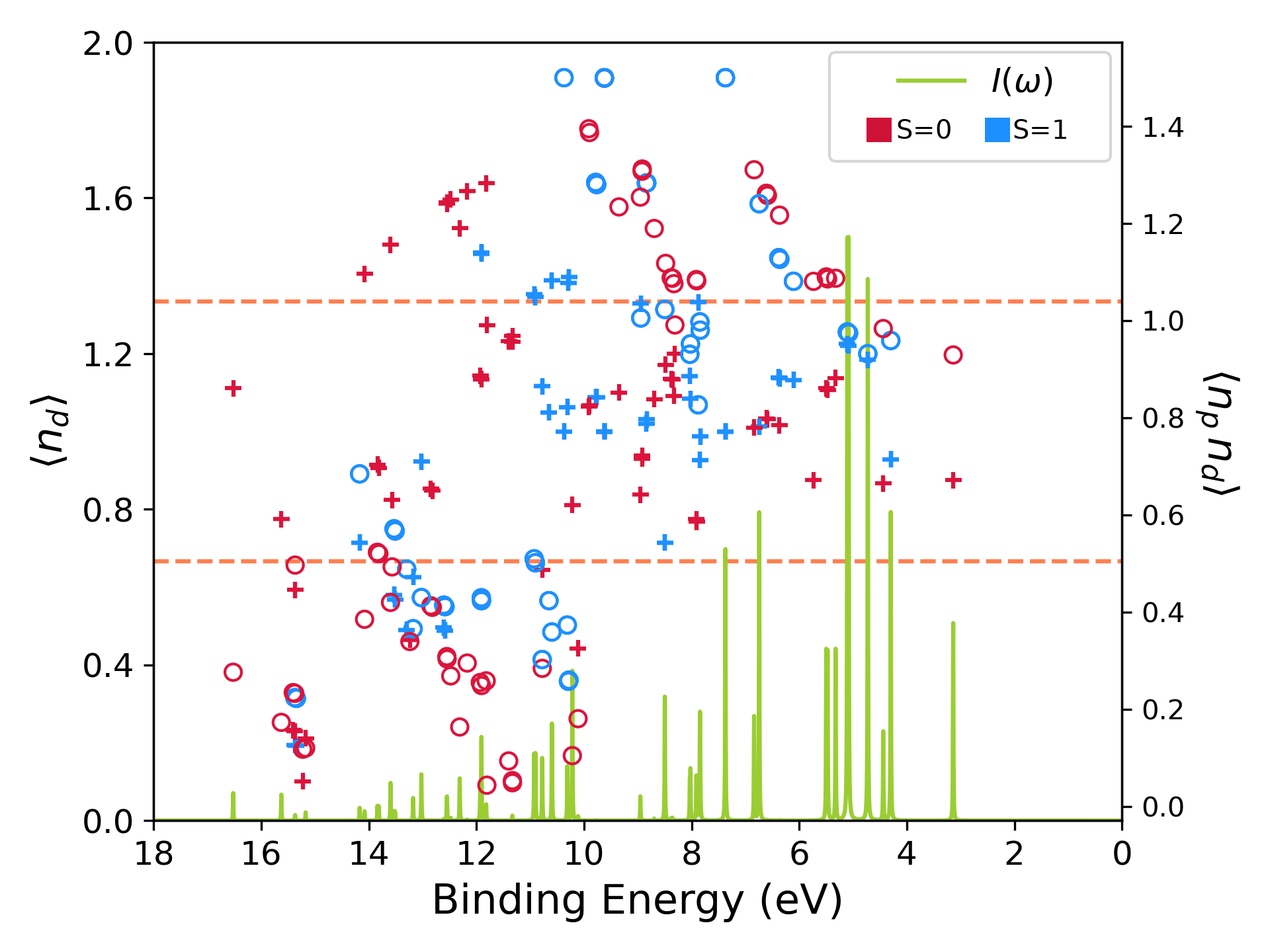}% Here is how to import EPS art
\caption{\label{fig:nddw} The crosses are the hole occupancy  
  of the $d$-states (left scale) while the circles are the interatomic charge correlator (right scale) in the possible final states of the photoemission process. The color of the symbols encode the singlet (red) or triplet (blue) character of the two-hole state. 
  Parameters and binding energies are defined as in Fig.~\ref{fig:agf2_vpes}. At the bottom we show the valence band spectrum on green to facilitate identification of the states.
}
\end{figure}

\subsubsection{Strongly bound Zhang-Rice state scenario}\label{sec:strongZR}
To illustrate the sensitivity of $U_{4d}$ and  $\Delta_{pd}$ to the choice of diagonal energies representing fluorine states we show the spectra with an {\em ad hoc} set of crystal fields. Namely 
we replace the values in Table~\ref{DFT_par_table} by 
$e_P^\nu=$ 1.5, -1.5, -1.5, 0.75 and 0.75 (all values in eV) for    $\nu=z^2, {x^2-y^2}, {xy}, {xz}$ and ${yz}$ respectively. Furthermore, 
following Refs.~\onlinecite{Ghijsen1988,Tjeng1990,Eskes1990} we neglect the crystal field on silver 
($\epsilon_d^\nu=0$). In this case we find that a good fit can be obtained with $U_{4d}=8.2$ eV and $\Delta_{pd}=3.5$ eV. The fit of the main shoulder is improved but there is still substantial spectral weight in the second shoulder which is not seen in the experiment.

\begin{figure}[tb]
  \includegraphics[scale=0.55]{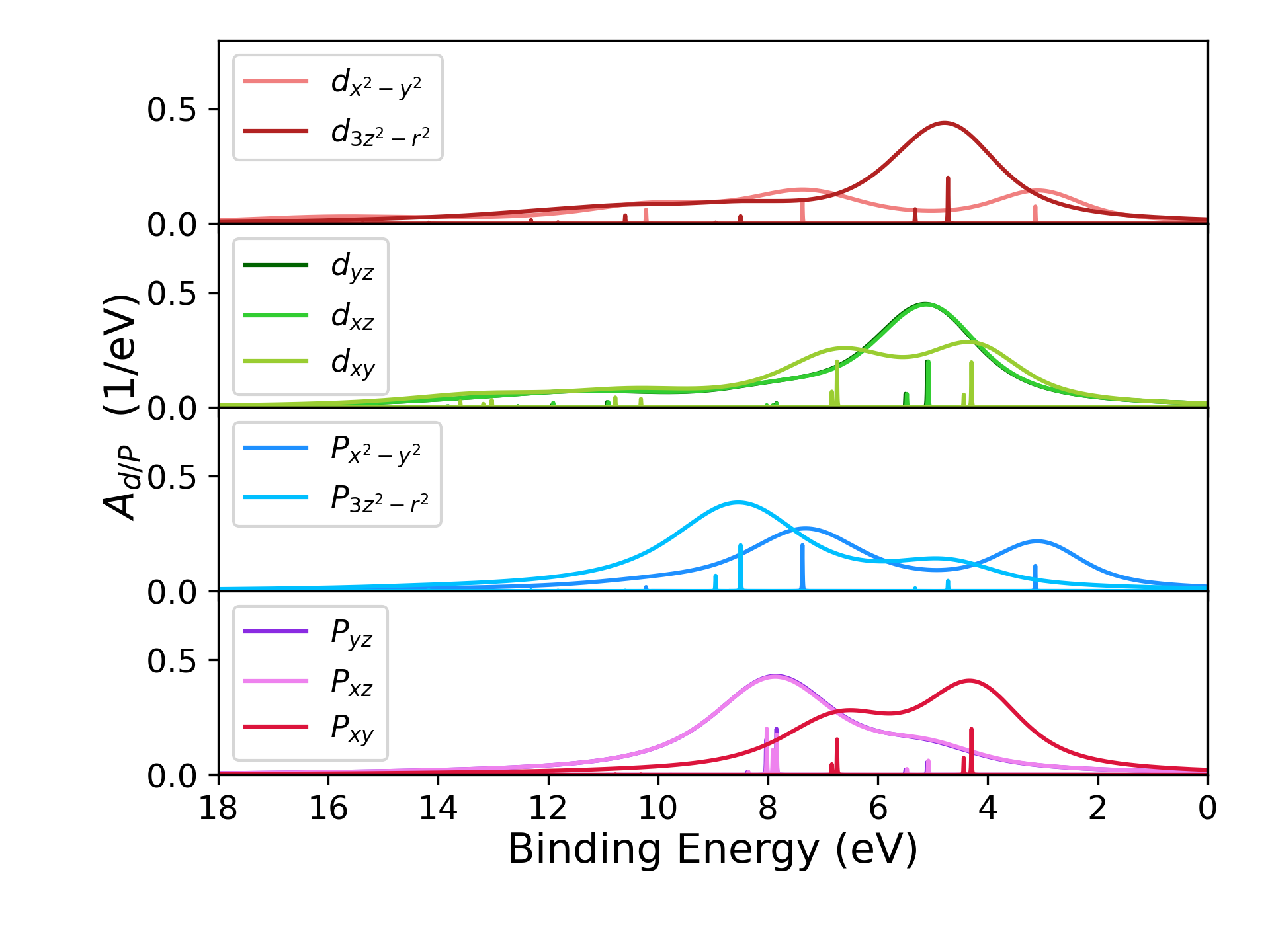}% Here is how to import EPS art
  \caption{\label{fig:Adpdw} One particle spectral function projected on the different orbitals symmetry of the cluster for the same parameters of Fig.~\ref{fig:agf2_vpes} }
\end{figure}

Now the ZR singlet is well separated from the rest of the spectra justifying the subsection title. From Figs.~\ref{fig:nddw} and \ref{fig:Adpdw}  one sees that the $P$ character of the ZR state is reinforced {with} respect to the previous case, increasing the resemblance {to} cuprates.
%The fundamental gap in this case is $E_{gap}=3.1$ eV. 

Since the present parameter set has larger  $U_{4d}$ and smaller $\Delta_{pd}$ the highest energy state in the removal spectra near 16.5 eV now has a significant component with the two holes in the $x^2-y^2$ orbital (cf. Figs.~\ref{fig:nddw} and \ref{fig:Adpdw})  instead of being dominated by $d^{10}\underline L^2$ configurations as in the {weakly bound ZR} case {presented in \ref{sec:weakZR}}. Unfortunately also this upper Hubbard band structure is not visible in the experiment, hampering a direct determination of $U_{4d}$.

In the previous case of small crystal fields parameters,  the spectra of the different symmetries  (except  $x^2-y^2$) resemble each other
(cf. Fig.~\ref{fig:Adpdw_cf_dft})
which leads to a clustering of peaks in the photoemission spectra with similar character of final states (cf. Fig.~\ref{fig:nddw_cf_dft}). For the larger crystal fields considered here, the peaks acquire a more heterogeneous character as witnessed by the expectation values in Fig.~\ref{fig:nddw}. The $d_{xy}$ spectral weight in Fig. \ref{fig:Adpdw} is shifted to smaller energies and differs from the other symmetries which is enough to substantially increase the agreement between theory and experiment.

%The energy of these states is larger than in cuprates both because of the larger $\Delta_{pd}$ and $U_{pp}$.
%The states above 12 eV binding energy have two holes on symmetrized ligand orbitals different from $x^2-y^2$ so are not accessible in photoemission.  

\begin{table}[t]
  \caption{\label{tab:dd} Energy of $dd$-transitions $x^2-y^2\rightarrow \nu$
    from Refs.~\onlinecite{Eskes1990} (Theory), \onlinecite{MorettiSala2011},\onlinecite{Bachar2021} (Experiment) and the present computations in the weakly bound (WB) and strongly bound (SB) ZR scenario. 
The first column labels the transitions by the final state $\nu$.
Since single crystals studies are not available, the $dd$ transitions of \ch{AgF2} have not yet been assigned to specific symmetries. We report the approximate location of the lower and higher energy feature which have been tentatively assigned to specific symmetries with the aid of the DFT computations in Ref.~\onlinecite{Bachar2021}.
For the experimental gap we report the peak in the optical conductivity from Refs.~\onlinecite{Falck1992} and~\onlinecite{Bachar2021}.
  }
\begin{ruledtabular}
  \begin{tabular}{c|cccc|cccc}
        &    \multicolumn{3}{c}{\ch{La2CuO4}} &    &\multicolumn{3}{c}{\ch{AgF2}} \\
        &    \multicolumn{3}{c}{ Theory   }    &Exp.& \multicolumn{2}{c}{Theory}&  Exp. \\
    Ref.    & \onlinecite{Eskes1990}    &\onlinecite{Eskes1990}&\onlinecite{Eskes1990}&\onlinecite{MorettiSala2011}&\multicolumn{2}{c}{This Work }& \onlinecite{Bachar2021} \\
         &     &       &     &      &WB ZR& SB ZR &      \\   
    \hline
 $z^2$  & 1.42& 1.28  & 1.20& 1.70&  0.96  &  1.51&  1.5  \\
 $x y$  & 1.53& 1.40  & 1.35& 1.80&  1.54  &  1.25&       \\   
 $x z$  & 1.60& 1.51  & 1.56& 2.12&  1.48  &  1.68&   2.4 \\
 $y z$  & 1.60& 1.51  & 1.56& 2.12&  1.50  &  1.70&       \\
 $E_{gap}$& 1.2 & 1.8   &  2.4&  2.2 &  3.6 & 3.1  &  3.4  \\
 $n_d$  & 0.60& 0.66  & 0.72&     &  0.80  &  0.67&       \\
\end{tabular}
\end{ruledtabular}
\end{table}

\subsection{ $dd$ excitations and gap}

A recent RIXS study\cite{Bachar2021} has found a set of $dd$ transitions
laying between 1.4~eV and 2.7~eV, very similar to the ones observed previously in cuprates. To explain the $dd$ transitions and a parallel optical study, similar cluster computations were performed
but using a model {that neglected} $U_{pp}$ and  $U_{pd}$. Fitting the optical and the RIXS spectra required using different parameters for optics and RIXS. 
{A 20\% increase of $T_{pd}$ with respect to the DFT value and a quite small value of the charge-transfer parameter, $\Delta=1.29$ eV, were required to match the $dd$ transitions.}
% Matching the $dd$ transitions required a 20\% increase of $T_{pd}$ respect to the DFT value and a quite small value of the charge-transfer parameter, $\Delta=1.29$ eV. 
On the other hand, optical excitations required a much larger value of the charge-transfer parameter,
$\Delta=2.8$ eV. It was argue{d} that the difference could be explained taking into account $U_{pd}\sim 1.5$ eV in an effective manner. 
Unfortunately the large number of free parameters does not allow {for} %to make
 a systematic fit of $U_{pd}${, hence} we fixed it to a value close to the one in   Ref.~\onlinecite{Bachar2021}.

From Table \ref{tab:dd} we see that 
the weakly bound ZR scenario underestimates the value of low energy $dd$ transitions while the strongly bound ZR scenario performs slightly better but underestimates the high energy part.
A similar problem occurs with the cluster computations of Ref.~\onlinecite{Eskes1990} shown on the same Table. The agreement can be improved taking into account interactions beyond the cluster as done in Ref.~\onlinecite{Hozoi2011}. For example within the cluster the one hole state is a mixture of $d^9$ and $d^{10}\underline L$ configurations with $x^2-y^2$ symmetry. Taking into account the interaction of these one-hole states beyond the cluster gives rise to the superexchange magnetic interaction which lowers the energy of the state by a quantity of the order of $J$ per Ag ion. This lowering will be nearly zero in the excited states since they are much more localized due to a smaller $T_{pd}^\nu$. Therefore, one should increase the excitation energy by a quantity of order\cite{Gawraczynski2019} $J=0.07$ eV in \ch{AgF2} and 0.1 eV in cuprates.
An accurate determination of  $dd$ transition{s} {of cuprates}  in quantum chemistry computations\cite{Hozoi2011,Huang2011} require the consideration of a large number of additional atoms {with} respect to what we considered here. So part {of} the disagreement found can be just a limitation of the present cluster computations and not 
%to be 
attributed to the parameters used.  
%For an even better agreement probably a larger cluster
%and the orbital dependence of $U_{pd}$ should be probably taken into account. 

The optical experiments of Ref.~\onlinecite{Bachar2021} in powder samples did not allow to determine the gap unambiguously. A peak value was reported at 3.4 eV but the onset of the non excitonic charge-transfer absorption was estimated around 2.2 eV. Such value, however, depends on assumptions about the powder average and the role of disorder on the measurements and should be taken with a grain of salt.
The fundamental gap for the parameters studied is shown in Table~\ref{tab:dd}. 
While the gap for the weakly bound ZR scenario is slightly too large, the uncertainties on the optical data do not allow to exclude this scenario on the basis of this comparison alone. 

\section{Comparison with the Hubbard $U$ interaction in the DFT+$U$ method}\label{sec:comp-with-hubb}
%Above 
{Until now} we have obtained estimates of the intra-orbital Hubbard $U_{4d}$ interaction in fluorides from spectroscopic experiments. Since the work of Anisimov and collaborators\cite{Anisimov1991}
it is customary in DFT computations of correlated systems to incorporate a Hubbard $U$ parameter (hereafter  $ U^*$) 
to cure some deficiencies of the particular functional in use. It is often assumed that the parameter of the DFT method should correspond to a physical repulsion {such as}  $ U^*=U_{4d}$. However, 
this equivalence is far from obvious. Indeed, for the exact functional  no correction is needed\cite{Cococcioni2005} and $ U^*=0$. Thus, it is clear that  $ U^*$ depends on the approximate functional in use{,} while  $U_{4d}$ is an observable spectroscopic quantity depending only on the material.

Setting $T_{pd}^\nu=0$ and taking the $d$ shell as an example one can quite generally define:
\begin{equation}
  \label{eq:defu}
 U_{4d}=E(d^{10})+E(^1A_{1g},d^8)-2E(d^9),
\end{equation}
where the notation emphasises that the $d^8$ state should be a singlet with both holes on the same orbital.  Assuming DFT  provides accurate total energies at integer occupations   it is reasonable to estimate the physical  $ U_{4d}$ using the above formula in a  DFT computation in which the $d$ shell occupancy  is constrained at  $n=8,9,10$ and the hybridization is suppressed. Indeed, this kind of computation was done even before the DFT+$U$ methodology was proposed\cite{McMahan1988,Hybertsen1990}.

The values of $E(d^{n})$ at three points ($n=8,9,10$) are enough to define a parabola.  
For common local or semilocal functionals, such as LDA or GGA, the energy as a function of $n$, taken now as a {\em continuous variable}, is expected to follow an approximate parabolic behavior. Instead, the energy computed with the exact functional is known to be piece-wise linear, coinciding with the parabolic behavior only at integer values of $n$.\cite{Cococcioni2005}
Cococcioni and De Gironcoli (CDG) proposed\cite{Cococcioni2005} to introduce $U^*$ in such a way that the piece-wise linear behavior is restored in the approximate DFT+$U$ functional. It is easy to see that if the energy computed with a given functional is a parabola {\em for all real} $n$ in the interval $8\leq n\leq 10$ then the CDG correction coincides also with the DFT estimate of the spectroscopic interaction using Eq.~(\ref{eq:defu}). {Therefore},
we can take $U^*= U_{4d}$ where,  the left{-}hand side is the Hubbard interaction in the DFT+$U$ method while the right{-}hand side should be understood as the DFT approximation to the spectroscopic interaction using Eq.~(\ref{eq:defu}).
This, of course, does not hold for an exact functional or for any functional which does not yield a perfect parabola in the full interval  $8\leq n\leq 10$. 

We have computed the Hubbard $U$ parameters both for Ag and F using the linear response approach of CDG as implemented in VASP software package.\cite{Kresse1996}\footnote{See also ``Calculate U for LSDA+U''- Vaspwiki available at \url{https://www.vasp.at/wiki/index.php/Calculate_U_for_LSDA\%2BU}}
The Monkhorst-Pack sampling of k-mesh was 6 x 6 x 7, and the energy cut-off for plane{-}wave basis was set as 520 eV. The PBEsol implementation of the exchange-correlation functional was used. 

We obtained $U^*(\ch{Ag})=5.31$eV and $U^*(\ch{F})=11.18$eV.
The value for Ag is smaller than  $U_{4d}=7\sim 8$ eV obtained from the spectroscopic data while the value for F is probably larger than the physical one. These disagreements are not uncommon. For example{,} for NiO Ref~\onlinecite{Cococcioni2005} finds $U^*(\ch{Ni})=4.6$eV{,} while the spectroscopic value\cite{VanElp1992} for antiparallel spin is $U_{3d}=10$ eV. This difference may be due to
inaccuracies of the DFT approximate energies at integer values (a problem that is simply not addressed by the CDG method) 
or to a non parabolic behavior of the energy in between. 
For closed shell ions (such as nominally \ch{F-}) overestimation of the Hubbard $U$ by the linear response method is common.\footnote{M. Cococcione, private communication.} 
A resolution of these issues is certainly very interesting and deserves further theoretical research but goes beyond our present scope.
Even if the physical and the corrective $U^*$ do not coincide, consistency of the method requires that the above value of $U^*(\ch{Ag})$ is used in DFT+$U$ computations with the PBEsol functional.

\section{Discussion and Conclusions}

We have presented a high energy spectroscopy study complemented with  cluster computations to determine the degree of correlations in silver fluorides and their similarities and/or differences with cuprates.

DFT computations predict {very similar}
$p-d$ hopping matrix elements %very similar
in the two compounds (c.f. Table.~\ref{tab:AgF2vsCuO}). Many important differences in electronic parameters can be traced back to the different size of the orbitals: \ch{Ag^{2+}} is larger than \ch{Cu^{2+}} while \ch{F-} is smaller than \ch{O^{2-}}.
The ion polarizability scales with the ionic volume{,} while the on-site Coulomb repulsion decreases with the size.
For the more localized fluorine $p$-orbitals we expect larger $U_{pp}$ and smaller $p-p$ hybridization and ligand polarizability in the silver fluoride. For 
$d$ orbitals the roles are interchanged and we expect smaller bare on-site interaction which is compensated by smaller screening by the ligands.

Our computations for closed shell AgF are in good agreement with the experiment. Using valence band photoemission and Auger spectroscopy we were able to estimate the Hubbard Coulomb repulsion among two holes on the same orbital to be $U_{4d}=7.7$~eV. As expected{,} this is smaller than the same quantity in cuprates which is partially compensated by the  screening effect.

For valence  band photoemission of \ch{AgF2} the situation is more complicated. The spectra 
shows a peak and a shoulder. This resembles the spectra observed in cuprates which, however, show also high binding energy satellites at a distance of $7\sim8$ eV from the main peak.  Unfortunately{,} such satellites {--} which in cuprates are related to $d^8$ states {--} are not seen here{, hampering} %so
a direct determination of   $U_{4d}$ in  \ch{AgF2}. %was not possible. 
Due to the large number of parameters and the small number of {observed} features %observed
a systematic fit was not possible. Thus, we considered two possible scenarios; a weakly bound and a strongly bound ZR singlet, that allowed to constrain the parameters $U_{4d}$ and $\Delta_{pd}$ and provide a reasonable range of possible values.
Table \ref{tab:AgF2vsCuO} compares the parameters found for \ch{AgF2} and \ch{CuO}.
%As pointed out previously,\cite{Gawraczynski2019,Bachar2021} $p-d$ hybridization is very similar to cuprates, while $p-p$ hybridization is smaller in the fluoride respect to the oxide.
In both {considered} scenarios{,} %considered
the charge transfer energy is somewhat larger and the Coulomb repulsion is smaller than in cuprates. However, in both scenarios the Coulomb repulsion dominates and the material can be classified as a charge transfer insulator according to the ZSA scheme.

The value found for  $\Delta_{pd}=1.2$ eV in AgF is considerably smaller than in \ch{AgF2}. In both materials  we defined $\Delta_{pd}$  as the energy cost of the ionic transition $d^{9}\rightarrow d^{10}\underline L$. The difference in $\Delta_{pd}$  can be understood considering the ionic potential.  The energy of the hole in the silver site is decreased the shorter the distance to the negatively charged fluorines. On the contrary, the energy of the hole on the ligand increases shortening the distance with the positively charged silver. Thus the increase of the charge transfer energy can be attributed in part to elementary electrostatic considerations and {to} the decrease of the Ag-F distance when going from AgF (2.47 $\si{\angstrom}$)\cite{Bottger1972} to \ch{AgF_2} (2.07 $\si{\angstrom}$)\cite{Fischer1971a}.

The weakly bound scenario used to explain the photoemission experiment in \ch{AgF2} predicts a small energy difference between a high-spin $d^8$ state and a low{-}spin ZR singlet.  
It is worth mentioning that in DFT doped holes on \ch{AgF2} planes 
form both low-spin\cite{Bandaru2021} and high-spin\cite{Jezierski2022} solutions depending on details such as the local strain and the environment. This sensitivity to details is compatible with a weakly bound ZR scenario. Unfortunately,
these computations neglect important correlation effects and should not be trusted to determine the first ionization state of the insulator for {reasons similar to the ones} discussed in Sec.~\ref{sec:comp-with-hubb}. Overall{,} the strongly bound ZR scenario gives a better match of the spectra and of $dd$-transitions{,} but the improvement is marginal {with} respect to the uncertainties involved so we cannot narrow more the range of parameters from what is shown in Table~\ref{tab:AgF2vsCuO}.

%With the present crystal field parameters, keeping  
% $U_{pp}$ allows to fit the experiment with a $\Delta_{pd}$ compatible with RIXS and optics as will be discussed next. 
%Overall the present scenario shows a better agreement with the experiment with regard to the charge gap, fit of the spectra and the energy of dd transitions. However the  

The puzzling  absence of $d^8$ satellites in the valence band photoemission spectrum can be due to overlap with other bands which strongly broaden these states. In cuprates it has been demonstrated that the photoemission intensity of {these} features can be strongly enhanced by tuning the photon energy to be resonant
with the core $3p$ threshold excitation\cite{Ghijsen1990} (74 eV) and the core $2p$ threshold\cite{Tjeng1991} (931 eV).  It would be highly desirable to perform similar experiments in \ch{AgF2}. A good candidate could be the  $4p$ edge (56 eV) which  shows\cite{Tjeng1990} a weak resonance in \ch{Ag2O}.

The photon energy used in this study (Al-K$\alpha$) makes the spectra to be dominated by the $d$-weight.
It would be very interesting to repeat the experiment with other light sources
which are sensitive to the $p$ spectral weight like He I and He II lines
[$\sigma(F 2p)/\sigma(Ag 4d)=0.84,0.34$ respectively].
This would allow to identify the structures predicted in the inset of Figs.~\ref{fig:agf2_vpes_dft},\ref{fig:agf2_vpes}. If the strongly bound ZR scenario is correct, a distinctive feature associated with this state should appear at low binding energy 
due to the high fluorine $p$ character of the removal states. Similar features are well known in cuprates\cite{Ghijsen1988}. 
This will unambiguously identify the character of the first removal state{,} which is also fundamental to determine the nature of possible hole doping that can be achieved in this compound. 

In general{,} the fit of the photoemission data in \ch{AgF2} requires a larger charge-transfer energy than {the one}
%what was
suggested by the analysis of $dd$ transitions in \ch{AgF2}{; on the other hand,} %while
analysis of optics data yield a similar value {to the one} found here. RIXS analysis of single crystal data would be highly desirable to identify the symmetry of the different $dd$ transitions and can serve to better constrain the crystal field splittings.

\begin{table}[t]
\caption{\label{tab:AgF2vsCuO} 
  Comparison of parameters between \ch{AgF2} (current work) and CuO (from Ref.~\onlinecite{Eskes1991}). %and \cite{Tjeng1989} ~\cite{Ghijsen1990}.
  All values are in eV. $ U_{4d}/U_{3d} $ is the intraorbital repulsion  $U(^1A_{1g})$. }
\begin{ruledtabular}
\begin{tabular}{c|c|c}
                  &\ch{AgF2}  &             \ch{CuO} \\\hline
$T_{pd}^{x^2-y^2}$  & 2.76            &      2.5    \\
  $T_{pp}$         &       0.11$^*$    &   1.0   \\
   $\Delta$        & $3.5 \sim 4.1 $          & 2.95   \\
  $ U_{4d}/U_{3d} $ & $7.2\sim 8.2$  & 8.8  \\  
  $U_{pp}$ &   6.0  &  -- %6.0$^{\dagger}$ %F0+.16F2
  \\
 $U_{pd}$ &    $1\sim 1.5 $    & $<1^\dagger$ \\
 $E_{gap}$ &  3.4   &  1.8\\
 $n_d$   &   $0.67\sim 0.80$  & 0.66 \\ 
\end{tabular}
$^*$For comparison for  AgF$_2$ we defined an effective $T_{pp}=[e_P(xz)+e_P(yz)]/2-e_P(x^2-y^2)$
with the crystal field parameters of Table \ref{DFT_par_table}.\\
$^\dagger$ $E_{gap}$ and $n_d$ are from Ref.~\onlinecite{Eskes1990} which did not include this parameter.
\end{ruledtabular}
%\vspace{1ex}
%     {\raggedright $^\dagger$ ($x^2-y^2$) symmetry.  \par}
\end{table}

 Our results confirm that \ch{AgF2} is a correlated insulator of charge-transfer type just as the insulating parent compounds of high-$T_c$ superconductors. 
 It  appears to have a  degree of covalency {similar to} cuprates and a somewhat smaller Coulomb repulsion on the transition metal ion. The repulsion on the ligand is expected to be larger than in cuprates. This, however, has a minor role in the low energy physics{,} as the probability of two holes to be on the same ligand is anyway small. The similarity with cuprates suggest{s} that \ch{AgF2} can become the corner stone of a new family of high-$T_c$ superconductors.

\begin{acknowledgments}
  We thank Matteo Cococcioni for useful discussions on the DFT+$U$ method.
  Research was carried out with the use of CePT infrastructure financed by the European Union - the European Regional Development Fund within the Operational Programme ``Innovative economy'' for 2007-2013 (POIG.02.02.00-14-024/08-00). The Polish authors are grateful to NCN for support (Maestro, 2017/26/A/ST5/00570).
  We acknowledge financial support from the Italian MIUR
  through Projects No. PRIN 2017Z8TS5B and 20207ZXT4Z and
  regione Lazio (L. R. 13/08) under project SIMAP. MNG
is supported by the Marie Sklodowska-Curie individual
fellowship Grant agreement SILVERPATH No: 893943.
We acknowledge the CINECA award under the ISCRA initiative Grants No. HP10C72OM1 and HP10BV0TBS,
for the availability of high performance computing resources and support.
Z.M. acknowledges the financial support of the Slovenian Research Agency
(research core funding No. P1-0045; Inorganic Chemistry and Technology).
\end{acknowledgments}

\appendix

%\section{}

\section{Treatment of Coulomb repulsion on fluorine}\label{sec:treatm-fluor-repuls}
Because $p$-orbitals in fluorine are more localized than in oxygen{,} intra-orbital repulsion may be more relevant. Therefore we treat it explicitly rather than in mean-field as {customarily} done in cuprates.

To reduce the Hilbert space in the cluster computations{,} we constructed symmetry adapted $P$-orbitals
which are linear combinations of $p$-orbitals with strong overlap with the $d$ orbitals. %Therefore we need 
{This requires} to write the atomic Coulomb operator on fluorine in the basis of $P$ orbitals. This defines the matrix elements 
$U^{PP}$ appearing in Eq.~(\ref{eq:h}). In the following we present the  derivation of this matrix.

For simplicity we take into account only the monopole part of the interaction on each fluorine atom. The 
Coulomb operator reads, 
$$H_{U_{pp}}=\frac{U_{pp}}2\sum_{r \alpha\alpha'\sigma\sigma'}  {\bar p}^\dagger_{r \alpha\sigma}  {\bar p}^\dagger_{r \alpha'\sigma'} {\bar p}_{r \alpha'\sigma'} {\bar p}_{r\alpha\sigma}$$ with ${\bar p}^\dagger_{r \alpha\sigma}$ creating an {\emph{electron}} in fluorine $r$ with orbital  and spin quantum number $\alpha\sigma$.
$r$ runs over the $N_F=6$ fluorine sites around the central Ag and $\alpha=x,y,z$.

The symmetry adapted orbitals are defined by
$$
{\bar P}_{m\sigma}^\dagger=\sum_{r\alpha}\phi_{m,r\alpha}{\bar p}_{r\alpha\sigma}^\dagger
$$
with $\phi_{m,r\alpha}$ a $18\times 18$ unitary matrix. 
The inverse transform reads: 
$$
{\bar p}_{r\alpha\sigma}^\dagger=\sum_{r\alpha}\phi_{r\alpha,m}^*{\bar P}_{m\sigma}^\dagger
$$
After {this} projection the interaction reads:
$$H_{U_{pp}}=\frac{U_{pp}}2\sum_{m_1m_2m_3m_4\sigma\sigma'} M^{m_1m_2}_{m_3m_4} {\bar P}^\dagger_{m_1\sigma}  {\bar P}^\dagger_{m_3\sigma'} {\bar P}_{m_4\sigma'} {\bar P}_{m_2\sigma}$$
%with ${\bar P}^\dagger_{r \alpha\sigma}$ creating an electron in fluoirine %$i$ with orbital  and spin quantum number $\alpha\sigma$
with
$$
M^{m_1m_2}_{m_3m_4}=\sum_{r\alpha\alpha'} \phi_{r\alpha,m_1}^* \phi_{r\alpha',m_3}^* \phi_{r\alpha',m_4} \phi_{r\alpha,m_2}.
$$
$M$ is a symmetric $18^2\times 18^2$ matrix with rows and columns labeled by
the pairs $m_1m_2$ and $m_3m_4$ with $m_i=1,..,3N_F$. In the Hilbert space of interest only a small portion of this matrix is needed{,} as discussed below. 

Transforming to the hole formalism of the main text ($\bar P^\dagger_{m\sigma} \rightarrow  P_{m\sigma}) $ one obtains, 
\begin{eqnarray}
  \label{eq:hupp}
  H_{U_{pp}}=&&E_{cs}+\sum_{m\sigma} \epsilon_{U_{PP}}^m   P^\dagger_{m\sigma} P_{m\sigma}\\
           &+&\frac{U_{pp}}2\sum_{m_1m_2m_3m_4\sigma\sigma'} M^{m_1m_2}_{m_3m_4} { P}^\dagger_{m_4\sigma}  { P}^\dagger_{m_2\sigma'} { P}_{m_1\sigma'} { P}_{m_3\sigma}  \nonumber
\end{eqnarray}
with the energy of the closed shell given by a direct and exchange contribution which can be put in the familiar form, 
$$
E_{cs}=U_{pp} \sum_{m_1m_2}2 \left(M^{m_1m_1}_{m_2m_2}-M^{m_1m_2}_{m_2m_1}\right)= U_{pp} N_F \frac{N_s (N_s-1)}2
$$
with $N_s=6$, the number of spin-orbitals in a $p$-shell.

After normal ordering one obtains single particle contributions defined by, 
$$
\epsilon_{U_{pp}}^m=-U_{pp}\sum_{m'} \left(2M_{mm}^{m'm'}-M_{mm'}^{m'm}\right)=-U_{pp}(N_s-1)
$$
This clearly represents the interaction of the removed electron with the other electrons in a given fluorine. Since this is independent of $m$ it can be incorporated in the definition of $\Delta$.
Following Ref.~\onlinecite{Bachar2021}{,} we can now separate the symmetry adapted orbitals into 5 orbitals with significant mixing with the central Ag and the rest. The former transform approximately as $d$ orbitals and are labeled accordingly as $m=z^2$, ${x^2-y^2}$, ${xy}$, $ {xz}$, and  ${yz}$. Neglecting irrelevant terms{,} the final form of the interaction reads as Eq.~(\ref{eq:h})
where now the sum is restricted to the symmetry adapted orbitals 
with $\nu_i\rightarrow m_i\sigma_i$ and 
$$
U^{(PP)}(m_1\sigma_1,m_2\sigma_2,m_3\sigma_3,m_4\sigma_4)=  \frac{U_{pp}}2 M_{m_2m_3}^{m_4m_1}\delta_{\sigma_1\sigma_4}\delta_{\sigma_2\sigma_3}
$$
\begin{figure}[tb]
\centering
\includegraphics[scale=0.7]{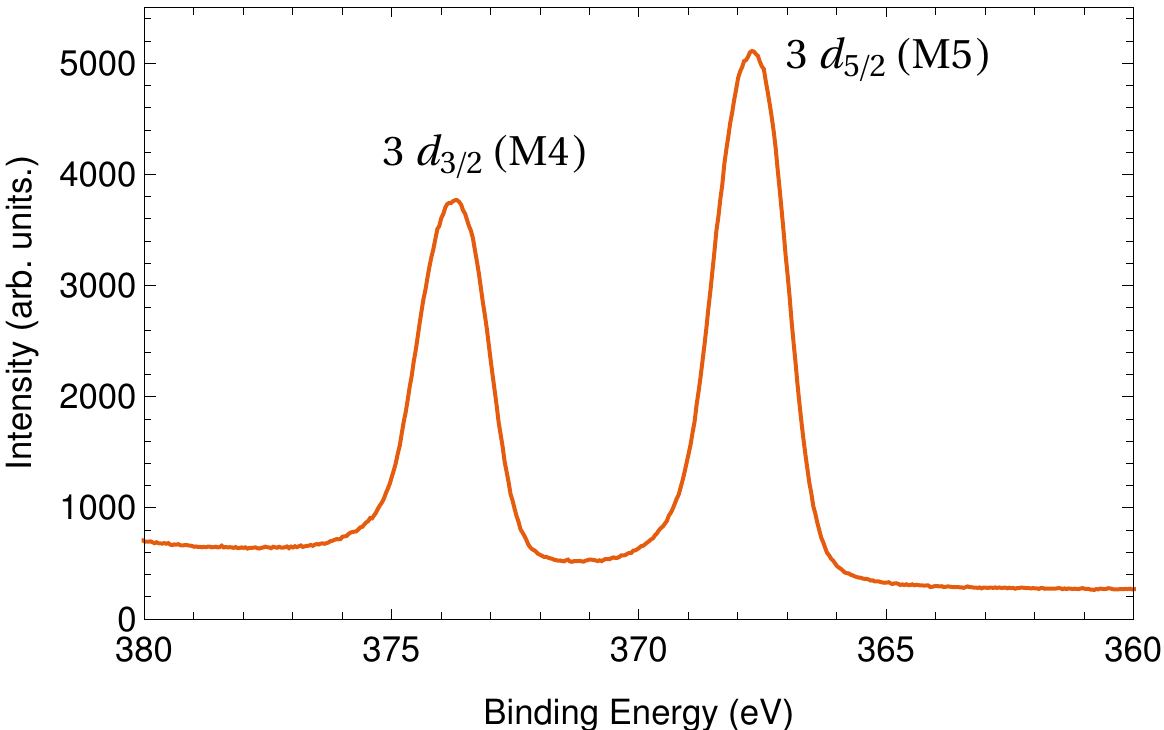}
\caption{\label{fig:AgFcore3d} The 3d-core hole spectra of AgF showing a $\Delta_{SO}=6$~eV splitting due to spin-orbit coupling.}
\end{figure}

 \section{Core-hole spectra and two-hole binding energies in Auger}\label{app:core-hole-spectra}

 In a two-step description of the Auger process{, the initial state consists in a core hole-state, followed by}
 %first a core hole-state is created which is the initial state for 
 the decay of a valence electron and the emission of a second valence electron ending in a state with two valence holes.   For AgF the initial core hole can be in a $3d_{3/2}$ ($M4$) or a $3d_{5/2}$ state ($M5$)  which are separated by the spin-orbit splitting $\Delta_{SO}=6.0$ eV as revealed by the core level spectra shown in 
 Fig.~\ref{fig:AgFcore3d}. Focusing on the case in which the initial core hole is in the $M5$ state the two-hole binding energy corresponding to a two-particle excitation $\nu$ is given by
 $E_b(\nu)=E(N-2,\nu)-E(N,0)=E(N-1,M5)-E(N,0)-KE$ where the kinetic energy of the photoejected electron is $KE=E(N-1,M5)-E(N-2,\nu)$. Here, $E(N,0)$ is the $N$-particle ground state. In Fig.~\ref{Auger_tot} {since} we 
 choose to plot the spectrum in terms of the $M5$ binding energy, the spectrum corresponding to the $M4$ initial state appears at smaller binding energy and shifted by the spin orbit coupling $\Delta_{SO}$.

%\bibliographystyle{prsty_no_etal}
%\bibliography{library}% Produces the bibliography via BibTeX.
%\bibliography{library,AgF2}% Produces the bibliography via BibTeX.

\end{document}